\documentclass[
 reprint,
 superscriptaddress,
preprintnumbers,
nofootinbib,
 amsmath,amssymb,
 aps,
prl,
]{revtex4-1}

\pdfoutput=1
\usepackage{amsmath}
\usepackage{amssymb}
\usepackage{amsthm}
\usepackage{MnSymbol}
\usepackage{nicefrac}
\usepackage{amsfonts}
\usepackage{dsfont}
\usepackage{natbib}
\usepackage{graphicx} 
\usepackage{hyperref}
\usepackage{slashed}
\newcommand{\bea}{\begin{eqnarray}}
\newcommand{\eea}{\end{eqnarray}}
\newcommand{\be}{\begin{equation}}
\newcommand{\ee}{\end{equation}}

\begin{document}

\title{
Safety in darkness: Higgs portal to simple Yukawa systems
}
 
 \author{Astrid Eichhorn}
   \email{eichhorn@sdu.dk}
\affiliation{CP3-Origins, University of Southern Denmark, Campusvej 55, DK-5230 Odense M, Denmark}
\affiliation{Institut f\"ur Theoretische
  Physik, Universit\"at Heidelberg, Philosophenweg 16, 69120
  Heidelberg, Germany}
\author{Martin Pauly}
\email{m.pauly@thphys.uni-heidelberg.de}
\affiliation{Institut f\"ur Theoretische
  Physik, Universit\"at Heidelberg, Philosophenweg 16, 69120
  Heidelberg, Germany}

\begin{abstract}
The nature of dark matter and the fundamental quantum structure of spacetime could be directly linked in the asymptotic-safety framework.
A toy model for the visible Higgs-Yukawa sector of the Standard Model, coupled to a dark sector through a portal coupling, provides a very first example for a model that simultaneously i) could become asymptotically safe at non-vanishing portal coupling, ii) could feature a strongly enhanced predictive power with calculable values for all interactions and thereby iii) give rise to calculable relations between the masses of the dark particles, their self-interactions, and the portal coupling.
\end{abstract}

\maketitle
Experimental searches for dark matter are ongoing, as of yet constraining parts of the vast space of dark-matter models, but with a discovery outstanding \cite{Wagner:2010mi,Hooper:2010mq,Ahnen:2016qkx,Aartsen:2016pfc,Akerib:2016vxi,Amole:2017dex,Agnese:2017jvy,Cui:2017nnn,Agnes:2018fwg,Aprile:2018dbl,Abdelhameed:2019hmk,Leane:2019xiy,Aguilar-Arevalo:2019wdi,Braine:2019fqb,Arnaud:2020svb,Andrianavalomahefa:2020ucg}. Thus, uncovering novel theoretical principles to guide the experimental searches is called for. At a first glance, the question, ``What is the fundamental nature of spacetime?'' appears to be an unrelated riddle. Yet, we will highlight that  a toy model for dark matter could be consistently embedded in a theory including quantum gravity within the asymptotic-safety paradigm. This could result in enhanced predictivity for the dark sector.\\
Asymptotic safety 
can provide an ultraviolet (UV) completion to an effective field theory (EFT), see \cite{Eichhorn:2018yfc} for a review. An EFT is characterized by an infinite number of  couplings and often features Landau poles in its Renormalization Group (RG) flow. 
Asymptotic safety corresponds to an interacting fixed point of the RG flow, i.e., an enhancement of the symmetry to quantum scale symmetry \cite{Wetterich:2019qzx}, providing a well-defined UV starting point for the RG flow. Such a UV completion is not available for arbitrary infrared (IR) values of the couplings. Instead, realizing quantum scale symmetry in the UV requires relations between couplings to hold. 
 Even power-counting renormalizable couplings -- generically free parameters in dark-matter models -- could become predictable in asymptotically safe dark-matter models with gravity \cite{Eichhorn:2017als,Reichert:2019car,Hamada:2020vnf}.  
For instance, there are indications that the Higgs portal coupling to a dark sector consisting of a single, uncharged dark scalar must vanish at the Planck scale in order to achieve asymptotic safety in the UV \cite{Eichhorn:2017als}. More generally, there are strong indications asymptotically safe quantum gravitational fluctuations flatten scalar potentials \cite{Zanusso:2009bs,Narain:2009fy,Narain:2009gb,Shaposhnikov:2009pv,Percacci:2015wwa,Labus:2015ska,Pawlowski:2018ixd,Wetterich:2019zdo,deBrito:2019umw,Eichhorn:2019dhg,Wetterich:2019rsn}. In \cite{Reichert:2019car,Hamada:2020vnf}, the resulting decoupling of scalar dark matter was circumvented by introducing an extended dark sector with a new gauge boson that regenerates the portal coupling during the RG flow. Here, we pursue two goals: \\
(i) We aim to realize a finite Higgs portal coupling directly as a consequence of asymptotic safety. \\
(ii) We explore the potential predictive power of asymptotic safety in such a portal model.

For quantum gravity, compelling indications for the asymptotically safe Reuter fixed point exist \cite{Reuter:1996cp,Reuter:2001ag,Lauscher:2002sq,Litim:2003vp,Codello:2008vh,Benedetti:2009rx,Falls:2013bv,Becker:2014qya,Christiansen:2015rva,Gies:2016con,Denz:2016qks,Eichhorn:2018ydy,Falls:2020qhj}, see \cite{Niedermaier:2006wt,Litim:2011cp,Percacci:2011fr,Reuter:2012id,Ashtekar:2014kba,Percacci:2017fkn,Eichhorn:2017egq,Pereira:2019dbn,Reuter:2019byg} for reviews and \cite{Eichhorn:2020mte,Reichert:2020mja} for introductory lectures. Among the open questions \cite{Donoghue:2019clr,Bonanno:2020bil}, the phenomenological viability is key -- as in any quantum-gravity approach. Consistency with the observed properties of matter already constitutes a test of asymptotically safe gravity  \cite{Eichhorn:2011pc,Dona:2013qba,Eichhorn:2017eht}. Indications exist, subject to systematic theoretical uncertainties, that quantum fluctuations of the Standard Model (SM) fields do not destroy asymptotic safety in gravity \cite{Dona:2013qba,Meibohm:2015twa,Biemans:2017zca,Alkofer:2018fxj}. Further, Euclidean calculations indicate that the values of several SM couplings might be
calculable quantities, fixed by requiring quantum scale symmetry in the UV \cite{Shaposhnikov:2009pv,Harst:2011zx,Eichhorn:2017ylw,Eichhorn:2017lry,Eichhorn:2017ylw}.  
These open the prospect of nontrivial observational consistency tests using existing measurements in particle physics.
Here, we move beyond consistency tests with existing observations and focus on the dark sector where predictions could be confronted with future experiments. To provide a proof-of-principle, we work in a toy model for a Higgs portal to dark matter  \cite{Silveira:1985rk,McDonald:1993ex,Burgess:2000yq,Bento:2001yk,McDonald:2001vt}, see \cite{Arcadi:2019lka} for a review.\\

\noindent\emph{Two simple Yukawa systems and a gravity-induced portal}\\
At the heart of the
decoupling-result in
\cite{Eichhorn:2017als} is shift symmetry for the dark scalar, forcing the portal to vanish \cite{Eichhorn:2017eht}. 
We go beyond existing decoupling results by (i) introducing a simple Yukawa system in the dark sector consisting of a dark scalar $\phi_d$ and a dark Dirac fermion $\psi_d$, see \cite{LopezHonorez:2012kv,Dupuis:2016fda,Matsumoto:2018acr}, (ii) incorporating the effect of non-minimal couplings. A finite fixed-point value for the dark Yukawa coupling breaks the shift symmetry in the dark scalar and gives rise to a finite non-minimal curvature coupling which generates a finite portal coupling.

We mimic the Higgs-top-sector of the  SM by a ``visible" scalar $\phi_v$ and a ``visible" Dirac fermion $\psi_v$. The single-scalar toy model  already captures the absence of massless Goldstone bosons for the SM Higgs and has been explored in the context of asymptotic safety without gravity in \cite{Gies:2009hq,Gies:2009sv,Vacca:2015nta}. 
 We study the following dynamics in a Euclidean setting
\be
\Gamma_k=\Gamma_k^{\rm visible} + \Gamma_k^{\rm dark} + \Gamma_k^{\rm portal}+ \Gamma_k^{\rm grav}, \label{eq:Gammak}
\ee
with
\be
\Gamma_k^{\rm grav}= \frac{-1}{16\pi G_N}\int d^4x\sqrt{g}\left(R- 2\bar{\Lambda} \right)  +  S_{\rm gauge-fixing},
\ee
\bea
\label{eqn:action_visible}
\Gamma_k^{\rm visible} &=& \int d^4x\sqrt{g}\Biggl(\frac{ Z_{\phi_v}}{2}g^{\mu\nu}\partial_{\mu}\phi_{v} \partial_{\nu}\phi_{v}+ \frac{\bar{m}_{v}^2}{2}\phi_{v}^2 + \frac{\lambda_{v}}{8}\phi_{v}^4\nonumber\\
&{}& \quad \quad- \xi_{v}\phi_{v}^2R + i { Z_{\psi_v}} \bar{\psi}_{v}\slashed{\nabla}\psi_{v} + i y_{v}\, \phi_{v} \bar{\psi}_{v}\psi_{v}\Biggr),\\
\Gamma_k^{\rm portal}&=& \int d^4x\sqrt{g}\, \frac{\lambda_{\rm HP}}{4}\phi_v^2\phi_d^2.
\eea
A crucial technical advancement is the inclusion of the non-minimal curvature couplings $\xi_{v(d)}$.
As we neglect the SM gauge group, there is an accidental exchange symmetry between dark and visible sectors. Thus, $\Gamma_k^{\rm dark}$ takes the same form as $\Gamma_k^{\rm visible}$, with the substitution $v \rightarrow d$, i.e., $\phi_v \rightarrow \phi_d$ etc. 
The model features two discrete $\mathbb{Z}_2$ symmetries. Under the $\mathbb{Z}_{2\, \rm dark (visible)}$, the dark (visible) sector transforms, while the visible (dark) sector is trivial. 
Spontaneous symmetry breaking of $\mathbb{Z}_{2\, \rm dark}$ is required for a massive dark fermion.
 In Eq.~\eqref{eq:Gammak}, $\Gamma_k$ is a scale-dependent effective action that allows us to set up a functional RG calculation in the Wilsonian spirit. 
 All fluctuations above the IR momentum cutoff $k$ are accounted for in the scale-dependent effective action $\Gamma_k$; fluctuations below $k$ remain to be integrated out. This is encoded in a flow equation for $\Gamma_k$ \cite{Wetterich:1992yh,Ellwanger:1993mw,Morris:1993qb}. It captures how the effective dynamics, parameterized by scale-dependent couplings, changes under a change in $k$, see, e.g., \cite{Pawlowski:2005xe,Gies:2006wv,Rosten:2010vm} for reviews and \cite{Reuter:2012id, Percacci:2017fkn,Reuter:2019byg} for reviews in the context of gravity. 
\\
At an RG fixed point, the scale dependence vanishes and quantum scale symmetry is realized. To connect a UV fixed point to physics in the IR, one follows an RG trajectory emanating out of the fixed point towards $k=0$. In this process, all quantum fluctuations are successively taken into account, until the physical limit $k=0$ has been reached. The number of relevant directions of the fixed point translates into the number of free parameters in the  description of the physics.
If the number of relevant parameters of the asymptotically safe fixed point is smaller than the number of perturbatively renormalizable couplings, the asymptotic-safety framework results in additional predictions compared to a perturbatively renormalizable model. In comparison to an EFT approach to dark matter, asymptotic safety also typically fixes higher-order couplings, which are free parameters of the EFT setting. \\

\begin{figure}[t!]
  \includegraphics[width=\linewidth]{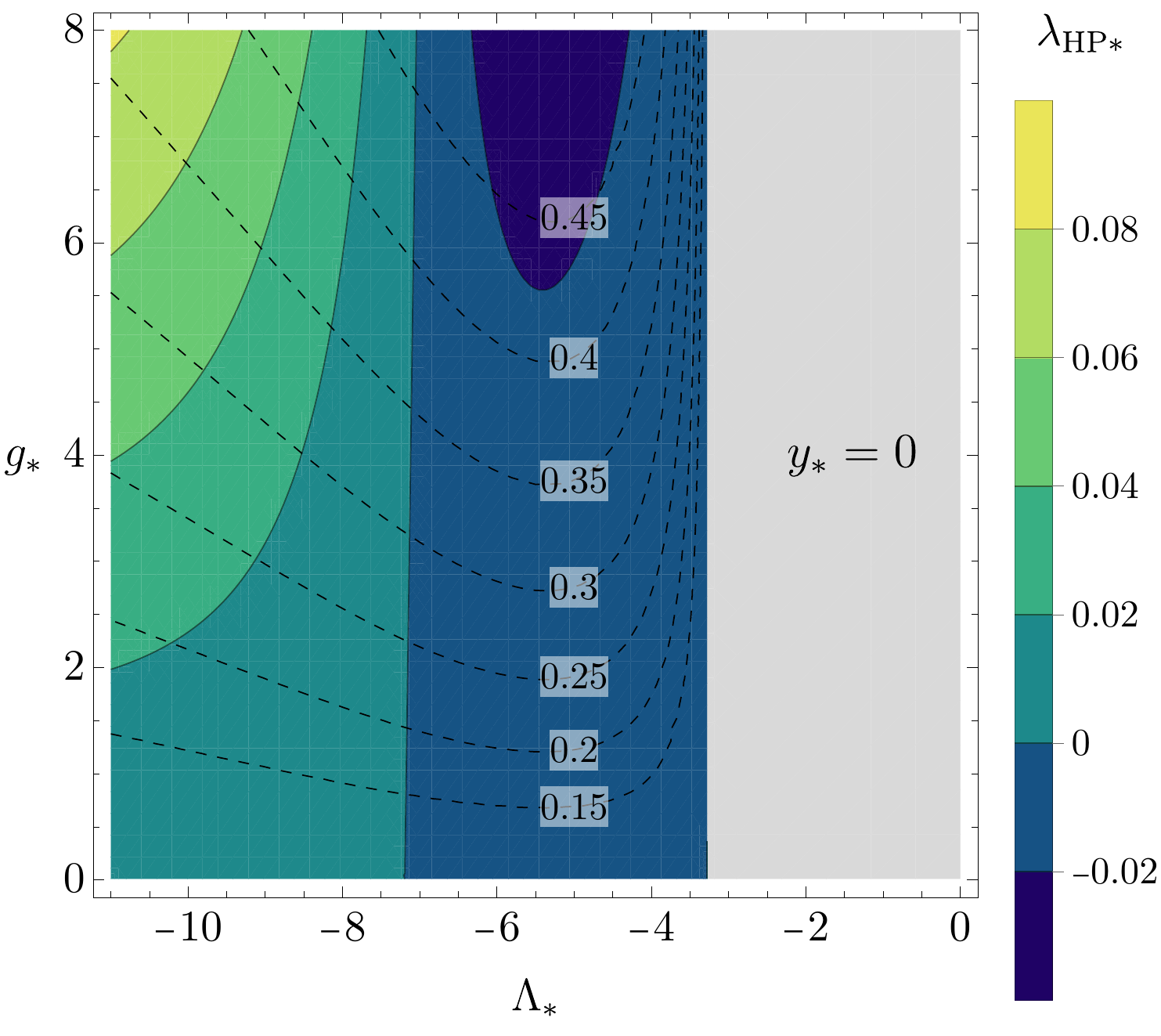}
  \caption{
    \label{fig:parameter_space}
We show the fixed-point values of the portal coupling $\lambda_{\rm HP\, \ast}$ (color coded) and the Yukawa coupling $y_{v(d)\, \ast}$ (dashed lines) as a function of the gravitational fixed-point values $\Lambda_\ast, g_\ast$. For $\Lambda_\ast \gtrsim -3.3$, no fixed point with non-vanishing Yukawa coupling exists \cite{Eichhorn:2017ylw}. For $\Lambda_\ast \lesssim -7.2$, the fixed-point value of the portal coupling turns positive. Even in the negative $\lambda_{\rm HP}$-region, the stability criterion $\lambda_{\rm HP}^2 - \lambda_{v}\lambda_d<0$ is fulfilled.
  }
\end{figure}

\noindent\emph{Mechanism for asymptotic safety in the portal coupling}\\
To search for quantum scale invariance, we work with the dimensionless counterparts of all couplings, defining $g= G_N\, k^2$, $\Lambda = \bar{\Lambda}\,k^{-2}$, $m_{v/d}^2 = \bar{m}_{v/d}^2 \,k^{-2}$.
Within the approximation defined by the truncation Eq.~\eqref{eq:Gammak}, the beta functions for the couplings $g$, $\Lambda$, $m_{v/d}$, $\lambda_{v/d}$, $\xi_{v/d}$, $y_{v/d}$ and $\lambda_{\rm HP}$ exhibit  multiple fixed points. We focus on the most predictive one. It features a nonvanishing portal coupling $\lambda_{\rm HP\,\ast}$ that generically translates into a nonvanishing portal coupling at $k=0$.
In the approximation defined by Eq.~\eqref{eq:Gammak}, the mechanism underlying this fixed point works as follows: 
Quantum fluctuations of gravity and matter fields generate an interacting fixed point for the gravitational couplings, $g$ and $\Lambda$. 
Even at finite values for $g$ and $\Lambda$, shift symmetry for the scalars protects the scalar potential in the absence of fermionic fluctuations \cite{Eichhorn:2017eht}, resulting in a flat potential with vanishing portal coupling, \cite{Eichhorn:2017als}.
 Non-vanishing Yukawa couplings in the dark and the visible sector break both shift symmetries. Their beta functions are given by
\be
\beta_{y_{v(d)}}=\frac{5\,y_{v(d)}^3}{16\pi^2} {\, - \,y_{v(d)}} f_y,
\ee
 with quantum gravitational fluctuations encoded in an anomalous dimension \cite{Eichhorn:2017ylw}, see also \cite{Zanusso:2009bs,Vacca:2010mj,Oda:2015sma,Eichhorn:2016esv,Hamada:2017rvn,deBrito:2019epw},
\bea
 f_y &=& -g\, \frac{96+\Lambda\left(-235+\Lambda(103+56\Lambda) \right)}{12 \pi\left(3+2 \Lambda(-5+4\Lambda) \right)^2}\nonumber\\
&{}&-g\, \xi_{ v(d)}\, \frac{6\left(18+8\Lambda+63 \xi_{ v(d)} - 168 \Lambda \xi_{ v(d)} \right)}{7\pi \left( 3-4\Lambda\right)^2}.
\eea
For $\Lambda$ negative enough and $|\xi|$ sufficiently small, $f_y$ is positive, cf.~Fig.~\ref{fig:parameter_space}, balancing out against the $y_{v(d)}^3$ term at an interacting fixed point 
\be
 y_{v(d)\, \ast} = \sqrt{\frac{16\pi^2\, f_y}{5}}
\ee
for the Yukawa couplings \cite{Eichhorn:2017ylw}, breaking shift symmetry in the UV.
The next step is the generation of a finite non-minimal coupling.
In the absence of the gravitational contribution the fixed-point value $\xi_{v(d)}=-1/12$ would be preferred \cite{Buchbinder:1992rb} at
finite values for the Yukawa couplings. The beta functions for the non-minimal couplings read,
\bea
\beta_{\xi_{v(d)}}&=& \frac{\left(4y_{v(d)}^2 + 3 \lambda_{v(d)}\right)(1+12 \xi_{v(d)})}{192\pi^2}+ \frac{\lambda_{\rm HP}(1+12\xi_{d(v)})}{192\pi^2}\nonumber\\
&-&\! \xi_{v(d)} f_{\xi}  \!+ \! g\, \xi_{v (d)}^2 \frac{72(21-8\Lambda)+972(5-8\Lambda)\xi_{v(d)}}{18\pi(3-4\Lambda)^2},
\label{eq:betaxi}
\eea
with
\be
f_{\xi}=g \frac{99+318 \Lambda -1464 \Lambda^2+1232 \Lambda^3-96 \Lambda^4}{18\pi(3-4 \Lambda)^2(1-2 \Lambda)^3},
\ee
 see, e.g., \cite{Narain:2009fy,Narain:2009gb,Percacci:2015wwa}. Gravity fluctuations together with Yukawa couplings result in a finite fixed-point value $\xi_{v(d)\, \ast}\neq -1/12$. 

The finite fixed-point value of the non-minimal couplings  induces a finite fixed-point value for the portal.
Its beta function reads
\bea
\beta_{\lambda_{\rm HP}} &=&  \frac{\lambda_{\rm HP}^2}{4\pi^2} + \frac{3}{16\pi^2}\lambda_{\rm HP} \left(\lambda_{v} + \lambda_d \right)  + \frac{\lambda_{\rm HP}}{4\pi^2}\left(y_v^2+y_d^2 \right)\nonumber\\
&{}&+ f_{\lambda}\, \lambda_{\rm HP}+ \beta_{\lambda_{\rm HP}}^{\rm ind}.
\eea
The first line is the one-loop result. The additional contributions arise from gravitational fluctuations, including the anomalous dimension
\bea
f_{\lambda} &=& g \Biggl(\frac{165-488 \Lambda+392 \Lambda^2-32 \Lambda^3}{6\pi\left(3+2 \Lambda(-5+4 \Lambda) \right)^2}+6\, \xi_{v}\,\xi_{d}\frac{108-96\Lambda}{\pi(3-4\Lambda)^2} \nonumber\\
&+& \frac{6(\xi_{v}+ \xi_{d})}{\pi(3-4\Lambda)^2}(18-8\Lambda) + \frac{6(\xi_{v}^2+ \xi_{d}^2)}{\pi(3-4\Lambda)^2}(45-72\Lambda) \Biggr).
\eea
As a key computational outcome, the non-minimal couplings generate the term,
\bea
\beta_{\lambda_{\rm HP}}^{\rm ind}
&=&  g^2\frac{160  \xi_{v}\xi_{d}}{(1-2 \Lambda)^3} + g^2 \frac{864 \xi_{v}\xi_{d} }{(3- 4\Lambda)^3}
+g^2\frac{82944}{(3-4\Lambda)^3}\xi_{v}^2\xi_{d}^2 \label{eq:ind1}\\
&+& g^2 \frac{27648}{(3-4\Lambda)^2}\xi_{v}^2\xi_{d}^2 + g^2 \frac{576(108-48\Lambda)}{5(3-4\Lambda)^3}\!\left(\!\xi_{v}^2\xi_{d}+ \xi_{d}^2\xi_{v} \!\right)\!.\nonumber
\eea
 The finite non-minimal couplings imply a  nonzero fixed-point value of the portal, $\lambda_{\rm HP\, \ast}\neq 0$. This is a central result of this paper, and provides the first example of a toy model for dark matter, in which a nonvanishing portal interaction is required by the microphysics.

The limitations of our results are rooted in the Euclidean setting and the necessity of truncating. The  effect of higher-order \emph{gravitational} couplings beyond our truncation amounts to a shift in the gravitational contributions e.g., $f_{\lambda}, f_y$, \cite{Oda:2015sma,Eichhorn:2017eht,deBrito:2019umw} 
as long as the gravitational higher-order couplings themselves feature a fixed point under the impact of matter, see, e.g., \cite{Oda:2015sma,Alkofer:2018fxj,Burger:2019upn}. Higher-order \emph{matter} couplings in scalar and fermion systems \cite{Eichhorn:2011pc,Eichhorn:2012va,Eichhorn:2016esv,Eichhorn:2017eht} contribute at sub-leading order to  the beta functions of the marginal couplings, at least at small numbers of matter fields. 
\\

\begin{figure}[!t]
  \includegraphics[width=\linewidth]{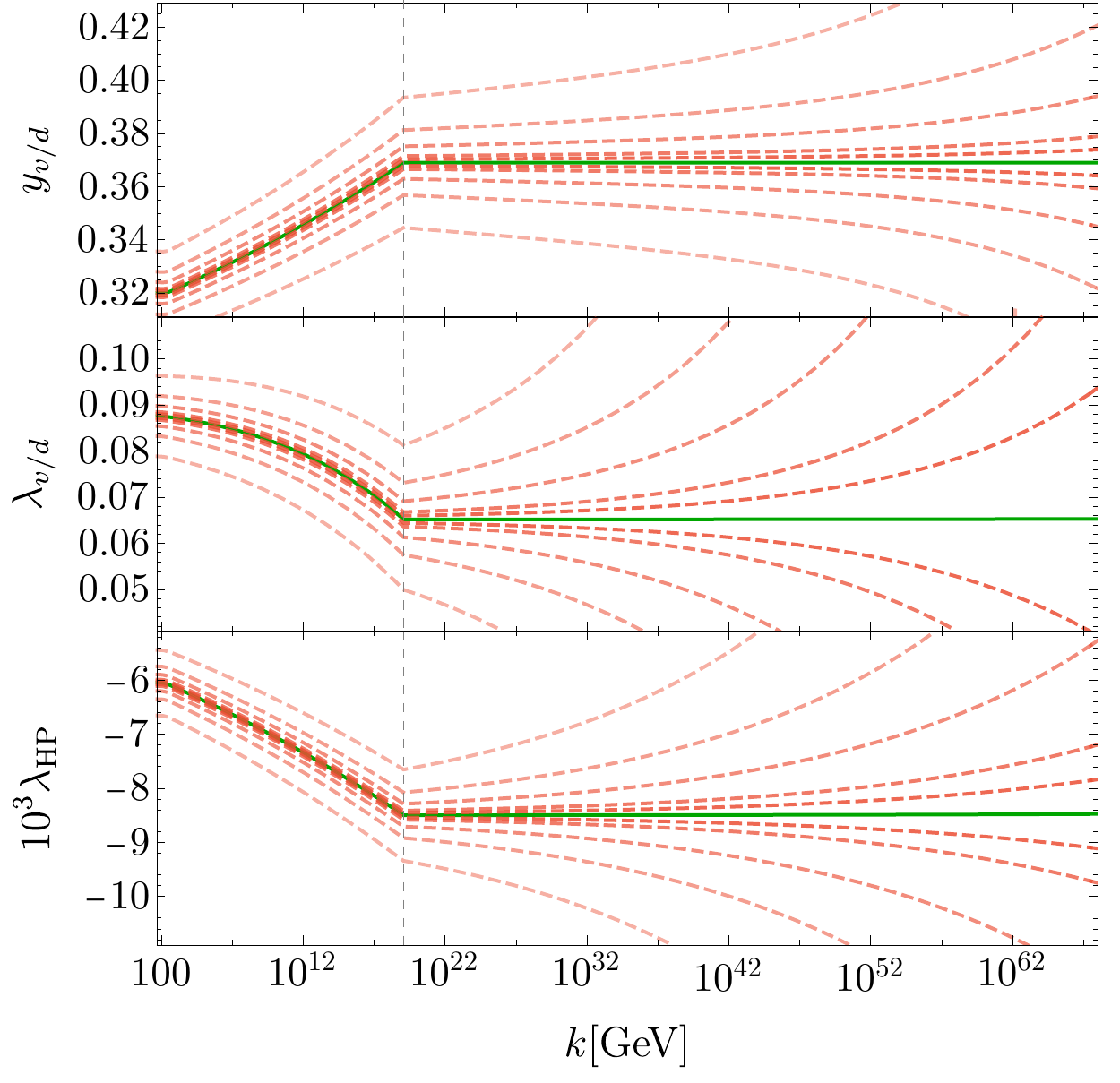}
  \caption{
    \label{fig:IRattractive}
We show trajectories for  one of the Yukawa couplings, one of the quartic couplings and the portal couplings as a function of $k$. For each case, the green, continuous trajectory is the unique trajectory that starts at the fixed point in the deep UV. These critical trajectories are IR attractive, such that even initial conditions that deviate from the critical trajectory in the UV are pulled towards it. At the Planck scale, indicated by the dashed vertical line, quantum-gravity fluctuations decouple dynamically. 
  }
\end{figure}

\begin{figure}[!t]
  \includegraphics[width=\linewidth]{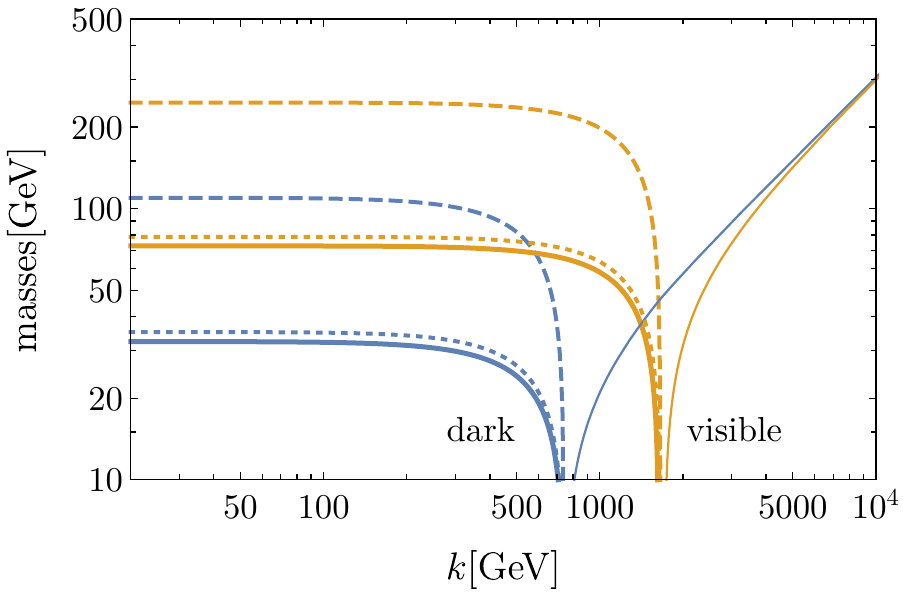}
  \caption{
    \label{fig:masses_over_k}
    The scalar mass $M_{V/D}$ (solid), vacuum expectation value $v_{v/d}$ (dashed) and fermion mass (dotted) $M_{\psi_{v/d}}$ as a function of the RG scale $k$. The vacuum expectation value in the visible sector is fixed such that $v_v \approx 246\, \text{GeV}$ in the IR. 
   The dark vacuum expectation value approaches $v_d \approx 109.5\, \text{GeV}$ in the above trajectories and is the free parameter of our toy model (see text). The dark sector is shown in blue and the visible sector in orange.
  }
\end{figure}

\noindent\emph{Predictivity from asymptotic safety}\\
To focus on our second goal and explore the predictivity of our toy-model scenario, we include the mass-parameters that are set to zero above
for readability. The full expressions are reported in Ref.~\cite{Eichhorn:2020sbo}.
We obtain fixed-point values  in the phenomenologically interesting region in Fig.~\ref{fig:parameter_space} 
\bea
g_{\ast}&=&4.55, \, \Lambda_{\ast}=-6.52,\label{eq:gravityFP}\\
 y_{v(d)\, \ast}&=&0.37, \, \xi_{v(d)\, \ast}=-2.7\cdot 10^{-2},\, \, m_{v(d)\, \ast}^2 = 1.6\cdot 10^{-3},\nonumber\\
  \lambda_{v(d)\, \ast}&=&6.5\cdot 10^{-2},\, \lambda_{\rm HP\, \ast}=-8.5\cdot 10^{-3}.\label{eq:matterFP}
\eea
Here, the gravitational beta functions as reported in \cite{Eichhorn:2017ylw} are used. We consider all SM fields ($4$ real scalars, $22.5$ Dirac fermions and $12$ gauge fields) plus the dark scalar and the dark Dirac fermion, resulting in a fixed point at negative microscopic cosmological constant, see also \cite{Dona:2013qba}. We neglected the impact of the mass and the non-minimal coupling on the gravitational fixed point, but confirmed that this only leads to a minor quantitative shift in our result.
The fixed point in Eq.~\eqref{eq:matterFP} not only features a finite portal coupling, but is also highly predictive: It is IR attractive in both Yukawa couplings, both non-minimal couplings and all quartic couplings 
\footnote{ 
This is encoded in the set of critical exponents, the real parts of which are given by
$ \theta_1 = 3.98,$  $\theta_2=1.92,$ $\theta_{3,4}=1.98,$  $\theta_{5,6}=-6.9\cdot10^{-3},$  $\theta_{7,8}=-7.1\cdot10^{-3},$
$\theta_{9}=-0.024,$  $\theta_{10,11}=-0.033.$
Every positive critical exponent signals one free parameter. The first two are related to the gravitational couplings, the next two are very close to the canonical dimension $\theta_{3/4,\text{can}} = 2$ for the mass parameters.
Small deviations $\frac{1}{9} \sum_{i=3}^{11} (\theta_i - \theta_{i,\text{can}})^2 \approx 0.02^2$ of the critical exponents in the matter sector from their canonical values $\theta_{i,\text{can}}$, with $\theta_{i, \text{can}}=0 \forall i > 4$, signal the near-perturbative nature of the fixed point.
Negative critical exponents also impose predictivity within effective asymptotic safety \cite{Held:2020kze}.
}.
The fixed-point requirement hence fixes the values of these couplings in the IR, cf.~Fig.~\ref{fig:IRattractive}.
The only free parameters at the fixed point 
in the matter sector are
the mass parameters of the two scalars.
The fixed-point properties indicate a near-perturbative nature, in line with research on symmetry identities \cite{Eichhorn:2018nda,Eichhorn:2018ydy,Eichhorn:2018akn}, extended truncations in the gravitational sector \cite{Falls:2013bv,Falls:2014tra,Falls:2017lst,Falls:2018ylp}, and perturbative computations \cite{Niedermaier:2009zz,Niedermaier:2010zz}. 
This signals the self-consistency of our approximation.
\\
\noindent\emph{Mass generation and symmetry breaking}\\
Starting from the UV fixed point, we follow the RG flow to the IR. We set the mass parameters for both scalars in the UV such that both $\mathbb{Z}_2$ symmetries are spontaneously broken in the IR and both fields acquire a vacuum expectation value (vev).
We assume that the visible scalar vev $v_v$ is known from measurements, and hold it fixed. The visible scalar mass is then predicted.
Such a mechanism has been proposed to predict the Higgs mass \cite{Shaposhnikov:2009pv}, see also \cite{Eichhorn:2017ylw,Pawlowski:2018ixd,Wetterich:2019zdo,Wetterich:2019rsn}.
 Then, the dark vev $v_d$ is the only free parameter in our model. 
After spontaneous symmetry breaking, the potential is most conveniently rewritten in the form
\be
V(\phi_v, \phi_d) \!=\! \sum_{i=v,d}\frac{\lambda_{i}}{8}\left(\phi_i^2 - v_i^2\right)^2
+ \frac{\lambda_{\rm HP}}{4}\left(\phi_v^2-v_v^2 \right)\left(\phi_d^2-v_d^2 \right).
\ee 
 This parametrization makes the symmetry breaking explicit and can directly be mapped to the potential terms in \eqref{eqn:action_visible}.
Once the RG scale $k$ drops below the mass of a mode, the corresponding mode automatically decouples from the  functional RG flow
 by virtue of non-trivial threshold functions.
 We focus on the case $\lambda_v v_v^2> \lambda_d v_d^2$.
The two massive scalars become superpositions of $\phi_v$ and $\phi_d$, with mass eigenvalues
\be
M_{V/D}^2 = \frac{1}{2}\left(\lambda_{v}v_v^2+ \lambda_d v_d^2 \pm \sqrt{(\lambda_v v_v^2-\lambda_d v_d^2)^2+ 4 \lambda_{\rm HP}^2 v_v^2\, v_d^2}\right),
\ee
The corresponding eigenstates feature a mixing angle 
\be
\tan 2  \alpha =\frac{ -2 v_v\, v_d\, \lambda_{\rm HP}}{\lambda_v v_v^2-\lambda_d v_d^2}.
\ee 
Fig.~\ref{fig:masses_over_k}  showcases the spontaneous symmetry breaking, followed by the automatic decoupling of fluctuations, where the  masses freeze out once $k$ drops below the physical mass scales. \\

\noindent\emph{Portal coupling versus dark scalar mass}

\begin{figure}
  \includegraphics[width=\linewidth]{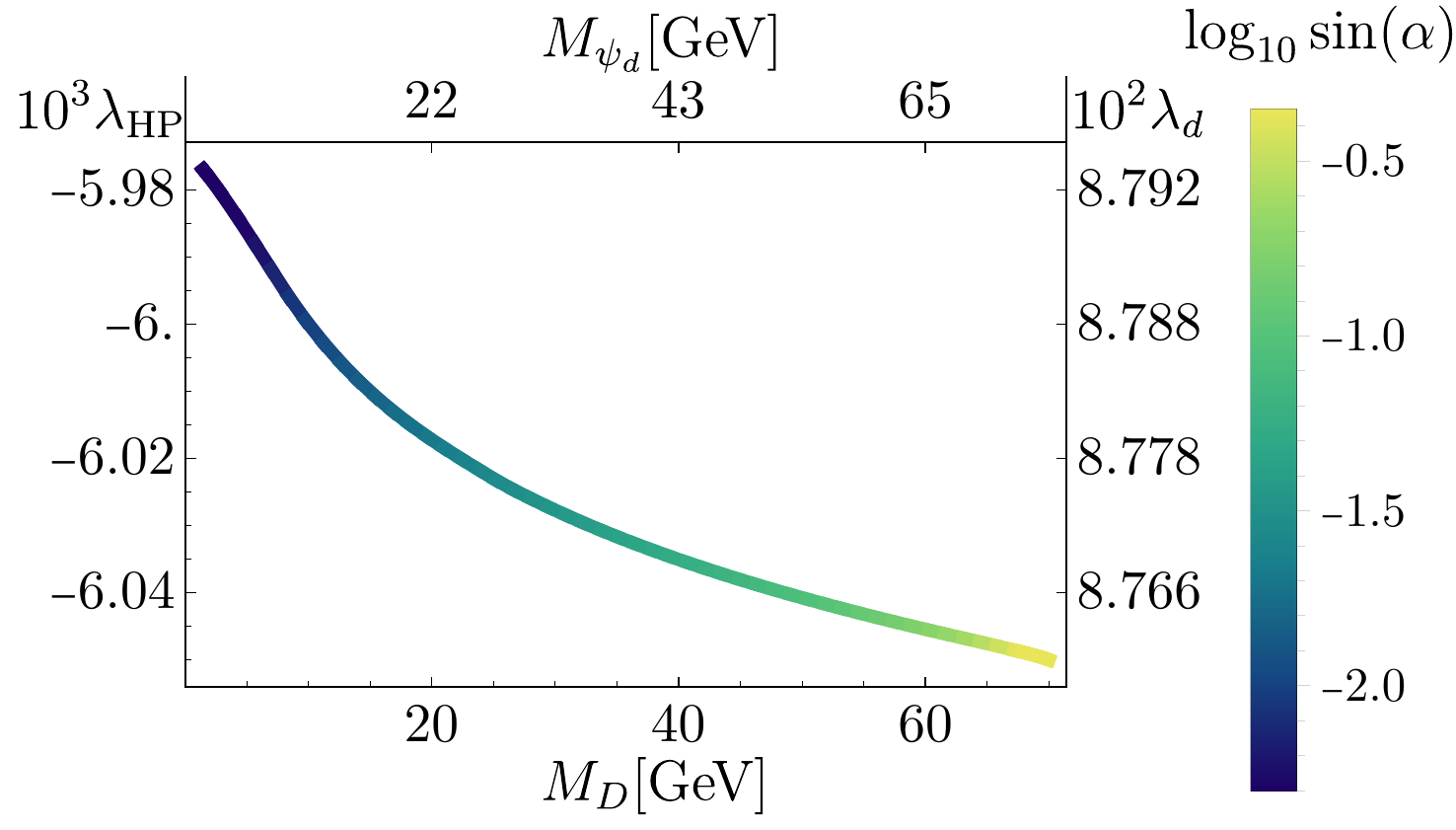}
  \caption{
    \label{fig:portalOverMass}
We show the relation between portal coupling and dark scalar mass that results from the asymptotically safe fixed point within the system of beta functions in our approximation. The mixing angle, dark self-interaction and dark-fermion mass  also become a function of the dark scalar mass. The corresponding scales at the right (top) are obtained by matching the corresponding values on the left (bottom), hence the scales are \textit{not} linear. The uncolored region is not compatible with the asymptotically safe fixed point in Eq.~\eqref{eq:matterFP}.
  }
\end{figure}

We map out the parameter space for dark matter in our asymptotically safe toy model by following RG trajectories from the UV to the IR. We vary the value of the dark scalar vev in the IR. As it corresponds to a relevant direction, a whole range of dark-scalar masses is compatible with a fixed point in the UV. In contrast, the other couplings are fixed uniquely as functions of the dark-scalar mass by the fixed-point requirement.
 The resulting relation between the dark-scalar mass and various other couplings is shown in Fig.~\ref{fig:portalOverMass}. It  illustrates the predictive power of asymptotic safety: The most important features are predictions for the portal coupling, dark fermion mass, dark-scalar self interaction, mixing angle and (not shown in the figure) non-minimal couplings as a function of the dark-scalar mass. The predicted portal coupling, $\lambda_{\rm HP} \approx -6\cdot 10^{-3}$ only varies mildly in response to changes in $M_D$.

In an extension of our toy-model to the SM -- if such an extension exists -- the $\mathbb{Z}_2$ symmetry for the dark scalar is broken spontaneously. Thus it decays to SM particles. The dark fermion might play the role of a stable dark matter candidate \cite{Baek:2011aa,Esch:2013rta,Bagherian:2014iia,Krnjaic:2015mbs}. 
The dark matter relic density is non-trivial to estimate, because the dark scalar and dark fermion have similar masses to each other and the Higgs resonance $M_D \lesssim M_{\psi_d} \sim M_V / 2$. We perform a rough estimate of the relic density. We take the IR values of the couplings in our toy model for the dark sector and use them in a calculation including SM fields. We implement a corresponding model in micrOMEGAs \cite{Belanger:2018ccd}, generating the relevant Feynman rules with LanHEP \cite{Semenov:2008jy}.
The dominant decay channel is $\psi_d \bar{\psi}_d \rightarrow \phi_d \phi_d$.
This yields a relic density in close vicinity to the observational value. We remind the reader that we are investigating a toy model. A more meaningful comparison with observational data is left to future work with a more realistic field content.\\

\noindent\emph{Outlook: Confronting asymptotically safe dark matter with observations and constraints}\\
We found indications that asymptotically safe quantum fluctuations of gravity and matter could  (i) generate a finite portal coupling and (ii) strongly enhance the predictive power of such a portal model.
In analogous studies of the Higgs-Yukawa sector, it has been found that similar fixed-point structures can tentatively be extended to a full SM gauge-Higgs-Yukawa sector \cite{Eichhorn:2017ylw,Eichhorn:2018whv,Alkofer:2020vtb}.
 Anticipating the extension of our results to a model including all SM degrees of freedom, this would suggest an asymptotically safe dark matter model with the following qualitative characteristics: The dark sector would be comprised of a scalar and a more massive Dirac fermion. The vacuum expectation values for the Higgs and the dark scalar would be the \emph{only} two free parameters in the scalar  Yukawa sector constrained experimentally by:
(i)  Higgs boson mass,
 (ii) quark masses,  (iii) observed dark-matter relic density \cite{Aghanim:2018eyx},   (iv) direct detection bounds \cite{Akerib:2016vxi,Agnese:2017jvy,Cui:2017nnn,Aprile:2018dbl}, (v) bounds on the invisible decay width of the Higgs \cite{OConnell:2006rsp,Aad:2015txa,Aad:2015pla,Dupuis:2016fda,Khachatryan:2016whc,Sirunyan:2018owy}.  Additionally, vacuum stability considerations in the early universe could potentially constrain the non-minimal couplings \cite{Figueroa:2017slm,Markkanen:2018pdo}.
These manifold non-trivial tests of a model that would be expected to feature only  two free parameters in the  scalar Yukawa sector  highlight the predictive power of the asymptotic-safety paradigm. 

This provides strong motivation for a concerted effort to develop a thorough quantitative understanding of the scenario we propose in the full SM
 while at the same time reducing systematic uncertainties due to the choice of truncation and spacetime signature. 
 Such a program might open up the perspective to meaningfully compare between an asymptotically safe gravity-matter model and experimental data in the future.
In this context, it is interesting that the value of the portal coupling in our toy model is of the right order of magnitude for a thermal relic and in the area of parameter space that is subject to intense observational scrutiny \cite{Cline:2013gha,Beniwal:2015sdl,Gaskins:2016cha,Arcadi:2017kky,Roszkowski:2017nbc,Athron:2018hpc} from \cite{Aprile:2015uzo,Aalbers:2016jon,Agnese:2016cpb,Akerib:2018lyp,Zhang:2018xdp}.

\acknowledgements{
We thank A.~Held for helpful discussions.
A.~E.~is supported by the DFG (Deutsche Forschungsgemeinschaft) under grant no.~Ei/1037-1, by a research grant (29405) from VILLUM FONDEN and by a visiting fellowship of the Perimeter Institute for Theoretical Physics. M.~P.~is supported by a scholarship of the German Academic Scholarship Foundation and is grateful to CP3-Origins, University of Southern Denmark, for extended hospitality.}

\bibliography{refs}

\begin{thebibliography}{129}%
\makeatletter
\providecommand \@ifxundefined [1]{%
 \@ifx{#1\undefined}
}%
\providecommand \@ifnum [1]{%
 \ifnum #1\expandafter \@firstoftwo
 \else \expandafter \@secondoftwo
 \fi
}%
\providecommand \@ifx [1]{%
 \ifx #1\expandafter \@firstoftwo
 \else \expandafter \@secondoftwo
 \fi
}%
\providecommand \natexlab [1]{#1}%
\providecommand \enquote  [1]{``#1''}%
\providecommand \bibnamefont  [1]{#1}%
\providecommand \bibfnamefont [1]{#1}%
\providecommand \citenamefont [1]{#1}%
\providecommand \href@noop [0]{\@secondoftwo}%
\providecommand \href [0]{\begingroup \@sanitize@url \@href}%
\providecommand \@href[1]{\@@startlink{#1}\@@href}%
\providecommand \@@href[1]{\endgroup#1\@@endlink}%
\providecommand \@sanitize@url [0]{\catcode `\\12\catcode `\$12\catcode
  `\&12\catcode `\#12\catcode `\^12\catcode `\_12\catcode `\%12\relax}%
\providecommand \@@startlink[1]{}%
\providecommand \@@endlink[0]{}%
\providecommand \url  [0]{\begingroup\@sanitize@url \@url }%
\providecommand \@url [1]{\endgroup\@href {#1}{\urlprefix }}%
\providecommand \urlprefix  [0]{URL }%
\providecommand \Eprint [0]{\href }%
\providecommand \doibase [0]{http://dx.doi.org/}%
\providecommand \selectlanguage [0]{\@gobble}%
\providecommand \bibinfo  [0]{\@secondoftwo}%
\providecommand \bibfield  [0]{\@secondoftwo}%
\providecommand \translation [1]{[#1]}%
\providecommand \BibitemOpen [0]{}%
\providecommand \bibitemStop [0]{}%
\providecommand \bibitemNoStop [0]{.\EOS\space}%
\providecommand \EOS [0]{\spacefactor3000\relax}%
\providecommand \BibitemShut  [1]{\csname bibitem#1\endcsname}%
\let\auto@bib@innerbib\@empty
\bibitem [{\citenamefont {Wagner}\ \emph {et~al.}(2010)\citenamefont {Wagner}
  \emph {et~al.}}]{Wagner:2010mi}%
  \BibitemOpen
  \bibfield  {author} {\bibinfo {author} {\bibfnamefont {A.}~\bibnamefont
  {Wagner}} \emph {et~al.} (\bibinfo {collaboration} {ADMX}),\ }\href {\doibase
  10.1103/PhysRevLett.105.171801} {\bibfield  {journal} {\bibinfo  {journal}
  {Phys. Rev. Lett.}\ }\textbf {\bibinfo {volume} {105}},\ \bibinfo {pages}
  {171801} (\bibinfo {year} {2010})},\ \Eprint {http://arxiv.org/abs/1007.3766}
  {arXiv:1007.3766 [hep-ex]} \BibitemShut {NoStop}%
\bibitem [{\citenamefont {Hooper}\ and\ \citenamefont
  {Goodenough}(2011)}]{Hooper:2010mq}%
  \BibitemOpen
  \bibfield  {author} {\bibinfo {author} {\bibfnamefont {D.}~\bibnamefont
  {Hooper}}\ and\ \bibinfo {author} {\bibfnamefont {L.}~\bibnamefont
  {Goodenough}},\ }\href {\doibase 10.1016/j.physletb.2011.02.029} {\bibfield
  {journal} {\bibinfo  {journal} {Phys. Lett. B}\ }\textbf {\bibinfo {volume}
  {697}},\ \bibinfo {pages} {412} (\bibinfo {year} {2011})},\ \Eprint
  {http://arxiv.org/abs/1010.2752} {arXiv:1010.2752 [hep-ph]} \BibitemShut
  {NoStop}%
\bibitem [{\citenamefont {Ahnen}\ \emph {et~al.}(2016)\citenamefont {Ahnen}
  \emph {et~al.}}]{Ahnen:2016qkx}%
  \BibitemOpen
  \bibfield  {author} {\bibinfo {author} {\bibfnamefont {M.}~\bibnamefont
  {Ahnen}} \emph {et~al.} (\bibinfo {collaboration} {MAGIC, Fermi-LAT}),\
  }\href {\doibase 10.1088/1475-7516/2016/02/039} {\bibfield  {journal}
  {\bibinfo  {journal} {JCAP}\ }\textbf {\bibinfo {volume} {02}},\ \bibinfo
  {pages} {039} (\bibinfo {year} {2016})},\ \Eprint
  {http://arxiv.org/abs/1601.06590} {arXiv:1601.06590 [astro-ph.HE]}
  \BibitemShut {NoStop}%
\bibitem [{\citenamefont {Aartsen}\ \emph {et~al.}(2016)\citenamefont {Aartsen}
  \emph {et~al.}}]{Aartsen:2016pfc}%
  \BibitemOpen
  \bibfield  {author} {\bibinfo {author} {\bibfnamefont {M.}~\bibnamefont
  {Aartsen}} \emph {et~al.} (\bibinfo {collaboration} {IceCube}),\ }\href
  {\doibase 10.1140/epjc/s10052-016-4375-3} {\bibfield  {journal} {\bibinfo
  {journal} {Eur. Phys. J. C}\ }\textbf {\bibinfo {volume} {76}},\ \bibinfo
  {pages} {531} (\bibinfo {year} {2016})},\ \Eprint
  {http://arxiv.org/abs/1606.00209} {arXiv:1606.00209 [astro-ph.HE]}
  \BibitemShut {NoStop}%
\bibitem [{\citenamefont {Akerib}\ \emph {et~al.}(2017)\citenamefont {Akerib}
  \emph {et~al.}}]{Akerib:2016vxi}%
  \BibitemOpen
  \bibfield  {author} {\bibinfo {author} {\bibfnamefont {D.}~\bibnamefont
  {Akerib}} \emph {et~al.} (\bibinfo {collaboration} {LUX}),\ }\href {\doibase
  10.1103/PhysRevLett.118.021303} {\bibfield  {journal} {\bibinfo  {journal}
  {Phys. Rev. Lett.}\ }\textbf {\bibinfo {volume} {118}},\ \bibinfo {pages}
  {021303} (\bibinfo {year} {2017})},\ \Eprint
  {http://arxiv.org/abs/1608.07648} {arXiv:1608.07648 [astro-ph.CO]}
  \BibitemShut {NoStop}%
\bibitem [{\citenamefont {Amole}\ \emph {et~al.}(2017)\citenamefont {Amole}
  \emph {et~al.}}]{Amole:2017dex}%
  \BibitemOpen
  \bibfield  {author} {\bibinfo {author} {\bibfnamefont {C.}~\bibnamefont
  {Amole}} \emph {et~al.} (\bibinfo {collaboration} {PICO}),\ }\href {\doibase
  10.1103/PhysRevLett.118.251301} {\bibfield  {journal} {\bibinfo  {journal}
  {Phys. Rev. Lett.}\ }\textbf {\bibinfo {volume} {118}},\ \bibinfo {pages}
  {251301} (\bibinfo {year} {2017})},\ \Eprint
  {http://arxiv.org/abs/1702.07666} {arXiv:1702.07666 [astro-ph.CO]}
  \BibitemShut {NoStop}%
\bibitem [{\citenamefont {Agnese}\ \emph {et~al.}(2018)\citenamefont {Agnese}
  \emph {et~al.}}]{Agnese:2017jvy}%
  \BibitemOpen
  \bibfield  {author} {\bibinfo {author} {\bibfnamefont {R.}~\bibnamefont
  {Agnese}} \emph {et~al.} (\bibinfo {collaboration} {SuperCDMS}),\ }\href
  {\doibase 10.1103/PhysRevD.97.022002} {\bibfield  {journal} {\bibinfo
  {journal} {Phys. Rev. D}\ }\textbf {\bibinfo {volume} {97}},\ \bibinfo
  {pages} {022002} (\bibinfo {year} {2018})},\ \Eprint
  {http://arxiv.org/abs/1707.01632} {arXiv:1707.01632 [astro-ph.CO]}
  \BibitemShut {NoStop}%
\bibitem [{\citenamefont {Cui}\ \emph {et~al.}(2017)\citenamefont {Cui} \emph
  {et~al.}}]{Cui:2017nnn}%
  \BibitemOpen
  \bibfield  {author} {\bibinfo {author} {\bibfnamefont {X.}~\bibnamefont
  {Cui}} \emph {et~al.} (\bibinfo {collaboration} {PandaX-II}),\ }\href
  {\doibase 10.1103/PhysRevLett.119.181302} {\bibfield  {journal} {\bibinfo
  {journal} {Phys. Rev. Lett.}\ }\textbf {\bibinfo {volume} {119}},\ \bibinfo
  {pages} {181302} (\bibinfo {year} {2017})},\ \Eprint
  {http://arxiv.org/abs/1708.06917} {arXiv:1708.06917 [astro-ph.CO]}
  \BibitemShut {NoStop}%
\bibitem [{\citenamefont {Agnes}\ \emph {et~al.}(2018)\citenamefont {Agnes}
  \emph {et~al.}}]{Agnes:2018fwg}%
  \BibitemOpen
  \bibfield  {author} {\bibinfo {author} {\bibfnamefont {P.}~\bibnamefont
  {Agnes}} \emph {et~al.} (\bibinfo {collaboration} {DarkSide}),\ }\href
  {\doibase 10.1103/PhysRevD.98.102006} {\bibfield  {journal} {\bibinfo
  {journal} {Phys. Rev. D}\ }\textbf {\bibinfo {volume} {98}},\ \bibinfo
  {pages} {102006} (\bibinfo {year} {2018})},\ \Eprint
  {http://arxiv.org/abs/1802.07198} {arXiv:1802.07198 [astro-ph.CO]}
  \BibitemShut {NoStop}%
\bibitem [{\citenamefont {Aprile}\ \emph {et~al.}(2018)\citenamefont {Aprile}
  \emph {et~al.}}]{Aprile:2018dbl}%
  \BibitemOpen
  \bibfield  {author} {\bibinfo {author} {\bibfnamefont {E.}~\bibnamefont
  {Aprile}} \emph {et~al.} (\bibinfo {collaboration} {XENON}),\ }\href
  {\doibase 10.1103/PhysRevLett.121.111302} {\bibfield  {journal} {\bibinfo
  {journal} {Phys. Rev. Lett.}\ }\textbf {\bibinfo {volume} {121}},\ \bibinfo
  {pages} {111302} (\bibinfo {year} {2018})},\ \Eprint
  {http://arxiv.org/abs/1805.12562} {arXiv:1805.12562 [astro-ph.CO]}
  \BibitemShut {NoStop}%
\bibitem [{\citenamefont {Abdelhameed}\ \emph {et~al.}(2019)\citenamefont
  {Abdelhameed} \emph {et~al.}}]{Abdelhameed:2019hmk}%
  \BibitemOpen
  \bibfield  {author} {\bibinfo {author} {\bibfnamefont {A.}~\bibnamefont
  {Abdelhameed}} \emph {et~al.} (\bibinfo {collaboration} {CRESST}),\ }\href
  {\doibase 10.1103/PhysRevD.100.102002} {\bibfield  {journal} {\bibinfo
  {journal} {Phys. Rev. D}\ }\textbf {\bibinfo {volume} {100}},\ \bibinfo
  {pages} {102002} (\bibinfo {year} {2019})},\ \Eprint
  {http://arxiv.org/abs/1904.00498} {arXiv:1904.00498 [astro-ph.CO]}
  \BibitemShut {NoStop}%
\bibitem [{\citenamefont {Leane}\ and\ \citenamefont
  {Slatyer}(2019)}]{Leane:2019xiy}%
  \BibitemOpen
  \bibfield  {author} {\bibinfo {author} {\bibfnamefont {R.~K.}\ \bibnamefont
  {Leane}}\ and\ \bibinfo {author} {\bibfnamefont {T.~R.}\ \bibnamefont
  {Slatyer}},\ }\href {\doibase 10.1103/PhysRevLett.123.241101} {\bibfield
  {journal} {\bibinfo  {journal} {Phys. Rev. Lett.}\ }\textbf {\bibinfo
  {volume} {123}},\ \bibinfo {pages} {241101} (\bibinfo {year} {2019})},\
  \Eprint {http://arxiv.org/abs/1904.08430} {arXiv:1904.08430 [astro-ph.HE]}
  \BibitemShut {NoStop}%
\bibitem [{\citenamefont {Aguilar-Arevalo}\ \emph {et~al.}(2019)\citenamefont
  {Aguilar-Arevalo} \emph {et~al.}}]{Aguilar-Arevalo:2019wdi}%
  \BibitemOpen
  \bibfield  {author} {\bibinfo {author} {\bibfnamefont {A.}~\bibnamefont
  {Aguilar-Arevalo}} \emph {et~al.} (\bibinfo {collaboration} {DAMIC}),\ }\href
  {\doibase 10.1103/PhysRevLett.123.181802} {\bibfield  {journal} {\bibinfo
  {journal} {Phys. Rev. Lett.}\ }\textbf {\bibinfo {volume} {123}},\ \bibinfo
  {pages} {181802} (\bibinfo {year} {2019})},\ \Eprint
  {http://arxiv.org/abs/1907.12628} {arXiv:1907.12628 [astro-ph.CO]}
  \BibitemShut {NoStop}%
\bibitem [{\citenamefont {Braine}\ \emph {et~al.}(2020)\citenamefont {Braine}
  \emph {et~al.}}]{Braine:2019fqb}%
  \BibitemOpen
  \bibfield  {author} {\bibinfo {author} {\bibfnamefont {T.}~\bibnamefont
  {Braine}} \emph {et~al.} (\bibinfo {collaboration} {ADMX}),\ }\href {\doibase
  10.1103/PhysRevLett.124.101303} {\bibfield  {journal} {\bibinfo  {journal}
  {Phys. Rev. Lett.}\ }\textbf {\bibinfo {volume} {124}},\ \bibinfo {pages}
  {101303} (\bibinfo {year} {2020})},\ \Eprint
  {http://arxiv.org/abs/1910.08638} {arXiv:1910.08638 [hep-ex]} \BibitemShut
  {NoStop}%
\bibitem [{\citenamefont {Arnaud}\ \emph {et~al.}(2020)\citenamefont {Arnaud}
  \emph {et~al.}}]{Arnaud:2020svb}%
  \BibitemOpen
  \bibfield  {author} {\bibinfo {author} {\bibfnamefont {Q.}~\bibnamefont
  {Arnaud}} \emph {et~al.} (\bibinfo {collaboration} {EDELWEISS}),\ }\href@noop
  {} {\  (\bibinfo {year} {2020})},\ \Eprint {http://arxiv.org/abs/2003.01046}
  {arXiv:2003.01046 [astro-ph.GA]} \BibitemShut {NoStop}%
\bibitem [{\citenamefont {Andrianavalomahefa}\ \emph
  {et~al.}(2020)\citenamefont {Andrianavalomahefa} \emph
  {et~al.}}]{Andrianavalomahefa:2020ucg}%
  \BibitemOpen
  \bibfield  {author} {\bibinfo {author} {\bibfnamefont {A.}~\bibnamefont
  {Andrianavalomahefa}} \emph {et~al.} (\bibinfo {collaboration} {FUNK
  Experiment}),\ }\href@noop {} {\  (\bibinfo {year} {2020})},\ \Eprint
  {http://arxiv.org/abs/2003.13144} {arXiv:2003.13144 [astro-ph.CO]}
  \BibitemShut {NoStop}%
\bibitem [{\citenamefont {Eichhorn}(2019)}]{Eichhorn:2018yfc}%
  \BibitemOpen
  \bibfield  {author} {\bibinfo {author} {\bibfnamefont {A.}~\bibnamefont
  {Eichhorn}},\ }\href {\doibase 10.3389/fspas.2018.00047} {\bibfield
  {journal} {\bibinfo  {journal} {Front. Astron. Space Sci.}\ }\textbf
  {\bibinfo {volume} {5}},\ \bibinfo {pages} {47} (\bibinfo {year} {2019})},\
  \Eprint {http://arxiv.org/abs/1810.07615} {arXiv:1810.07615 [hep-th]}
  \BibitemShut {NoStop}%
\bibitem [{\citenamefont {Wetterich}(2019{\natexlab{a}})}]{Wetterich:2019qzx}%
  \BibitemOpen
  \bibfield  {author} {\bibinfo {author} {\bibfnamefont {C.}~\bibnamefont
  {Wetterich}},\ }\href@noop {} {\  (\bibinfo {year} {2019}{\natexlab{a}})},\
  \Eprint {http://arxiv.org/abs/1901.04741} {arXiv:1901.04741 [hep-th]}
  \BibitemShut {NoStop}%
\bibitem [{\citenamefont {Eichhorn}\ \emph
  {et~al.}(2018{\natexlab{a}})\citenamefont {Eichhorn}, \citenamefont {Hamada},
  \citenamefont {Lumma},\ and\ \citenamefont {Yamada}}]{Eichhorn:2017als}%
  \BibitemOpen
  \bibfield  {author} {\bibinfo {author} {\bibfnamefont {A.}~\bibnamefont
  {Eichhorn}}, \bibinfo {author} {\bibfnamefont {Y.}~\bibnamefont {Hamada}},
  \bibinfo {author} {\bibfnamefont {J.}~\bibnamefont {Lumma}}, \ and\ \bibinfo
  {author} {\bibfnamefont {M.}~\bibnamefont {Yamada}},\ }\href {\doibase
  10.1103/PhysRevD.97.086004} {\bibfield  {journal} {\bibinfo  {journal} {Phys.
  Rev.}\ }\textbf {\bibinfo {volume} {D97}},\ \bibinfo {pages} {086004}
  (\bibinfo {year} {2018}{\natexlab{a}})},\ \Eprint
  {http://arxiv.org/abs/1712.00319} {arXiv:1712.00319 [hep-th]} \BibitemShut
  {NoStop}%
\bibitem [{\citenamefont {Reichert}\ and\ \citenamefont
  {Smirnov}(2020)}]{Reichert:2019car}%
  \BibitemOpen
  \bibfield  {author} {\bibinfo {author} {\bibfnamefont {M.}~\bibnamefont
  {Reichert}}\ and\ \bibinfo {author} {\bibfnamefont {J.}~\bibnamefont
  {Smirnov}},\ }\href {\doibase 10.1103/PhysRevD.101.063015} {\bibfield
  {journal} {\bibinfo  {journal} {Phys. Rev. D}\ }\textbf {\bibinfo {volume}
  {101}},\ \bibinfo {pages} {063015} (\bibinfo {year} {2020})},\ \Eprint
  {http://arxiv.org/abs/1911.00012} {arXiv:1911.00012 [hep-ph]} \BibitemShut
  {NoStop}%
\bibitem [{\citenamefont {Hamada}\ \emph {et~al.}(2020)\citenamefont {Hamada},
  \citenamefont {Tsumura},\ and\ \citenamefont {Yamada}}]{Hamada:2020vnf}%
  \BibitemOpen
  \bibfield  {author} {\bibinfo {author} {\bibfnamefont {Y.}~\bibnamefont
  {Hamada}}, \bibinfo {author} {\bibfnamefont {K.}~\bibnamefont {Tsumura}}, \
  and\ \bibinfo {author} {\bibfnamefont {M.}~\bibnamefont {Yamada}},\ }\href
  {\doibase 10.1140/epjc/s10052-020-7929-3} {\bibfield  {journal} {\bibinfo
  {journal} {Eur. Phys. J. C}\ }\textbf {\bibinfo {volume} {80}},\ \bibinfo
  {pages} {368} (\bibinfo {year} {2020})},\ \Eprint
  {http://arxiv.org/abs/2002.03666} {arXiv:2002.03666 [hep-ph]} \BibitemShut
  {NoStop}%
\bibitem [{\citenamefont {Zanusso}\ \emph {et~al.}(2010)\citenamefont
  {Zanusso}, \citenamefont {Zambelli}, \citenamefont {Vacca},\ and\
  \citenamefont {Percacci}}]{Zanusso:2009bs}%
  \BibitemOpen
  \bibfield  {author} {\bibinfo {author} {\bibfnamefont {O.}~\bibnamefont
  {Zanusso}}, \bibinfo {author} {\bibfnamefont {L.}~\bibnamefont {Zambelli}},
  \bibinfo {author} {\bibfnamefont {G.~P.}\ \bibnamefont {Vacca}}, \ and\
  \bibinfo {author} {\bibfnamefont {R.}~\bibnamefont {Percacci}},\ }\href
  {\doibase 10.1016/j.physletb.2010.04.043} {\bibfield  {journal} {\bibinfo
  {journal} {Phys. Lett.}\ }\textbf {\bibinfo {volume} {B689}},\ \bibinfo
  {pages} {90} (\bibinfo {year} {2010})},\ \Eprint
  {http://arxiv.org/abs/0904.0938} {arXiv:0904.0938 [hep-th]} \BibitemShut
  {NoStop}%
\bibitem [{\citenamefont {Narain}\ and\ \citenamefont
  {Percacci}(2010)}]{Narain:2009fy}%
  \BibitemOpen
  \bibfield  {author} {\bibinfo {author} {\bibfnamefont {G.}~\bibnamefont
  {Narain}}\ and\ \bibinfo {author} {\bibfnamefont {R.}~\bibnamefont
  {Percacci}},\ }\href {\doibase 10.1088/0264-9381/27/7/075001} {\bibfield
  {journal} {\bibinfo  {journal} {Class. Quant. Grav.}\ }\textbf {\bibinfo
  {volume} {27}},\ \bibinfo {pages} {075001} (\bibinfo {year} {2010})},\
  \Eprint {http://arxiv.org/abs/0911.0386} {arXiv:0911.0386 [hep-th]}
  \BibitemShut {NoStop}%
\bibitem [{\citenamefont {Narain}\ and\ \citenamefont
  {Rahmede}(2010)}]{Narain:2009gb}%
  \BibitemOpen
  \bibfield  {author} {\bibinfo {author} {\bibfnamefont {G.}~\bibnamefont
  {Narain}}\ and\ \bibinfo {author} {\bibfnamefont {C.}~\bibnamefont
  {Rahmede}},\ }\href {\doibase 10.1088/0264-9381/27/7/075002} {\bibfield
  {journal} {\bibinfo  {journal} {Class. Quant. Grav.}\ }\textbf {\bibinfo
  {volume} {27}},\ \bibinfo {pages} {075002} (\bibinfo {year} {2010})},\
  \Eprint {http://arxiv.org/abs/0911.0394} {arXiv:0911.0394 [hep-th]}
  \BibitemShut {NoStop}%
\bibitem [{\citenamefont {Shaposhnikov}\ and\ \citenamefont
  {Wetterich}(2010)}]{Shaposhnikov:2009pv}%
  \BibitemOpen
  \bibfield  {author} {\bibinfo {author} {\bibfnamefont {M.}~\bibnamefont
  {Shaposhnikov}}\ and\ \bibinfo {author} {\bibfnamefont {C.}~\bibnamefont
  {Wetterich}},\ }\href {\doibase 10.1016/j.physletb.2009.12.022} {\bibfield
  {journal} {\bibinfo  {journal} {Phys. Lett.}\ }\textbf {\bibinfo {volume}
  {B683}},\ \bibinfo {pages} {196} (\bibinfo {year} {2010})},\ \Eprint
  {http://arxiv.org/abs/0912.0208} {arXiv:0912.0208 [hep-th]} \BibitemShut
  {NoStop}%
\bibitem [{\citenamefont {Percacci}\ and\ \citenamefont
  {Vacca}(2015)}]{Percacci:2015wwa}%
  \BibitemOpen
  \bibfield  {author} {\bibinfo {author} {\bibfnamefont {R.}~\bibnamefont
  {Percacci}}\ and\ \bibinfo {author} {\bibfnamefont {G.~P.}\ \bibnamefont
  {Vacca}},\ }\href {\doibase 10.1140/epjc/s10052-015-3410-0} {\bibfield
  {journal} {\bibinfo  {journal} {Eur.\ Phys.\ J.\ C}\ }\textbf {\bibinfo
  {volume} {75}},\ \bibinfo {pages} {188} (\bibinfo {year} {2015})},\ \Eprint
  {http://arxiv.org/abs/1501.00888} {arXiv:1501.00888 [hep-th]} \BibitemShut
  {NoStop}%
\bibitem [{\citenamefont {Labus}\ \emph {et~al.}(2016)\citenamefont {Labus},
  \citenamefont {Percacci},\ and\ \citenamefont {Vacca}}]{Labus:2015ska}%
  \BibitemOpen
  \bibfield  {author} {\bibinfo {author} {\bibfnamefont {P.}~\bibnamefont
  {Labus}}, \bibinfo {author} {\bibfnamefont {R.}~\bibnamefont {Percacci}}, \
  and\ \bibinfo {author} {\bibfnamefont {G.~P.}\ \bibnamefont {Vacca}},\ }\href
  {\doibase 10.1016/j.physletb.2015.12.022} {\bibfield  {journal} {\bibinfo
  {journal} {Phys. Lett.}\ }\textbf {\bibinfo {volume} {B753}},\ \bibinfo
  {pages} {274} (\bibinfo {year} {2016})},\ \Eprint
  {http://arxiv.org/abs/1505.05393} {arXiv:1505.05393 [hep-th]} \BibitemShut
  {NoStop}%
\bibitem [{\citenamefont {Pawlowski}\ \emph {et~al.}(2019)\citenamefont
  {Pawlowski}, \citenamefont {Reichert}, \citenamefont {Wetterich},\ and\
  \citenamefont {Yamada}}]{Pawlowski:2018ixd}%
  \BibitemOpen
  \bibfield  {author} {\bibinfo {author} {\bibfnamefont {J.~M.}\ \bibnamefont
  {Pawlowski}}, \bibinfo {author} {\bibfnamefont {M.}~\bibnamefont {Reichert}},
  \bibinfo {author} {\bibfnamefont {C.}~\bibnamefont {Wetterich}}, \ and\
  \bibinfo {author} {\bibfnamefont {M.}~\bibnamefont {Yamada}},\ }\href
  {\doibase 10.1103/PhysRevD.99.086010} {\bibfield  {journal} {\bibinfo
  {journal} {Phys. Rev.}\ }\textbf {\bibinfo {volume} {D99}},\ \bibinfo {pages}
  {086010} (\bibinfo {year} {2019})},\ \Eprint
  {http://arxiv.org/abs/1811.11706} {arXiv:1811.11706 [hep-th]} \BibitemShut
  {NoStop}%
\bibitem [{\citenamefont {Wetterich}\ and\ \citenamefont
  {Yamada}(2019)}]{Wetterich:2019zdo}%
  \BibitemOpen
  \bibfield  {author} {\bibinfo {author} {\bibfnamefont {C.}~\bibnamefont
  {Wetterich}}\ and\ \bibinfo {author} {\bibfnamefont {M.}~\bibnamefont
  {Yamada}},\ }\href {\doibase 10.1103/PhysRevD.100.066017} {\bibfield
  {journal} {\bibinfo  {journal} {Phys. Rev.}\ }\textbf {\bibinfo {volume}
  {D100}},\ \bibinfo {pages} {066017} (\bibinfo {year} {2019})},\ \Eprint
  {http://arxiv.org/abs/1906.01721} {arXiv:1906.01721 [hep-th]} \BibitemShut
  {NoStop}%
\bibitem [{\citenamefont {De~Brito}\ \emph
  {et~al.}(2019{\natexlab{a}})\citenamefont {De~Brito}, \citenamefont
  {Eichhorn},\ and\ \citenamefont {Pereira}}]{deBrito:2019umw}%
  \BibitemOpen
  \bibfield  {author} {\bibinfo {author} {\bibfnamefont {G.~P.}\ \bibnamefont
  {De~Brito}}, \bibinfo {author} {\bibfnamefont {A.}~\bibnamefont {Eichhorn}},
  \ and\ \bibinfo {author} {\bibfnamefont {A.~D.}\ \bibnamefont {Pereira}},\
  }\href {\doibase 10.1007/JHEP09(2019)100} {\bibfield  {journal} {\bibinfo
  {journal} {JHEP}\ }\textbf {\bibinfo {volume} {09}},\ \bibinfo {pages} {100}
  (\bibinfo {year} {2019}{\natexlab{a}})},\ \Eprint
  {http://arxiv.org/abs/1907.11173} {arXiv:1907.11173 [hep-th]} \BibitemShut
  {NoStop}%
\bibitem [{\citenamefont {Eichhorn}\ \emph
  {et~al.}(2019{\natexlab{a}})\citenamefont {Eichhorn}, \citenamefont {Held},\
  and\ \citenamefont {Wetterich}}]{Eichhorn:2019dhg}%
  \BibitemOpen
  \bibfield  {author} {\bibinfo {author} {\bibfnamefont {A.}~\bibnamefont
  {Eichhorn}}, \bibinfo {author} {\bibfnamefont {A.}~\bibnamefont {Held}}, \
  and\ \bibinfo {author} {\bibfnamefont {C.}~\bibnamefont {Wetterich}},\
  }\href@noop {} {\  (\bibinfo {year} {2019}{\natexlab{a}})},\ \Eprint
  {http://arxiv.org/abs/1909.07318} {arXiv:1909.07318 [hep-th]} \BibitemShut
  {NoStop}%
\bibitem [{\citenamefont {Wetterich}(2019{\natexlab{b}})}]{Wetterich:2019rsn}%
  \BibitemOpen
  \bibfield  {author} {\bibinfo {author} {\bibfnamefont {C.}~\bibnamefont
  {Wetterich}},\ }\href@noop {} {\  (\bibinfo {year} {2019}{\natexlab{b}})},\
  \Eprint {http://arxiv.org/abs/1911.06100} {arXiv:1911.06100 [hep-th]}
  \BibitemShut {NoStop}%
\bibitem [{\citenamefont {Reuter}(1998)}]{Reuter:1996cp}%
  \BibitemOpen
  \bibfield  {author} {\bibinfo {author} {\bibfnamefont {M.}~\bibnamefont
  {Reuter}},\ }\href {\doibase 10.1103/PhysRevD.57.971} {\bibfield  {journal}
  {\bibinfo  {journal} {Phys. Rev.}\ }\textbf {\bibinfo {volume} {D57}},\
  \bibinfo {pages} {971} (\bibinfo {year} {1998})},\ \Eprint
  {http://arxiv.org/abs/hep-th/9605030} {arXiv:hep-th/9605030} \BibitemShut
  {NoStop}%
\bibitem [{\citenamefont {Reuter}\ and\ \citenamefont
  {Saueressig}(2002)}]{Reuter:2001ag}%
  \BibitemOpen
  \bibfield  {author} {\bibinfo {author} {\bibfnamefont {M.}~\bibnamefont
  {Reuter}}\ and\ \bibinfo {author} {\bibfnamefont {F.}~\bibnamefont
  {Saueressig}},\ }\href {\doibase 10.1103/PhysRevD.65.065016} {\bibfield
  {journal} {\bibinfo  {journal} {Phys. Rev.}\ }\textbf {\bibinfo {volume}
  {D65}},\ \bibinfo {pages} {065016} (\bibinfo {year} {2002})},\ \Eprint
  {http://arxiv.org/abs/hep-th/0110054} {arXiv:hep-th/0110054 [hep-th]}
  \BibitemShut {NoStop}%
\bibitem [{\citenamefont {Lauscher}\ and\ \citenamefont
  {Reuter}(2002)}]{Lauscher:2002sq}%
  \BibitemOpen
  \bibfield  {author} {\bibinfo {author} {\bibfnamefont {O.}~\bibnamefont
  {Lauscher}}\ and\ \bibinfo {author} {\bibfnamefont {M.}~\bibnamefont
  {Reuter}},\ }\href {\doibase 10.1103/PhysRevD.66.025026} {\bibfield
  {journal} {\bibinfo  {journal} {Phys. Rev.}\ }\textbf {\bibinfo {volume}
  {D66}},\ \bibinfo {pages} {025026} (\bibinfo {year} {2002})},\ \Eprint
  {http://arxiv.org/abs/hep-th/0205062} {arXiv:hep-th/0205062} \BibitemShut
  {NoStop}%
\bibitem [{\citenamefont {Litim}(2004)}]{Litim:2003vp}%
  \BibitemOpen
  \bibfield  {author} {\bibinfo {author} {\bibfnamefont {D.~F.}\ \bibnamefont
  {Litim}},\ }\href {\doibase 10.1103/PhysRevLett.92.201301} {\bibfield
  {journal} {\bibinfo  {journal} {Phys.Rev.Lett.}\ }\textbf {\bibinfo {volume}
  {92}},\ \bibinfo {pages} {201301} (\bibinfo {year} {2004})},\ \Eprint
  {http://arxiv.org/abs/hep-th/0312114} {arXiv:hep-th/0312114 [hep-th]}
  \BibitemShut {NoStop}%
\bibitem [{\citenamefont {Codello}\ \emph {et~al.}(2009)\citenamefont
  {Codello}, \citenamefont {Percacci},\ and\ \citenamefont
  {Rahmede}}]{Codello:2008vh}%
  \BibitemOpen
  \bibfield  {author} {\bibinfo {author} {\bibfnamefont {A.}~\bibnamefont
  {Codello}}, \bibinfo {author} {\bibfnamefont {R.}~\bibnamefont {Percacci}}, \
  and\ \bibinfo {author} {\bibfnamefont {C.}~\bibnamefont {Rahmede}},\ }\href
  {\doibase 10.1016/j.aop.2008.08.008} {\bibfield  {journal} {\bibinfo
  {journal} {Annals Phys.}\ }\textbf {\bibinfo {volume} {324}},\ \bibinfo
  {pages} {414} (\bibinfo {year} {2009})},\ \Eprint
  {http://arxiv.org/abs/0805.2909} {arXiv:0805.2909 [hep-th]} \BibitemShut
  {NoStop}%
\bibitem [{\citenamefont {Benedetti}\ \emph {et~al.}(2009)\citenamefont
  {Benedetti}, \citenamefont {Machado},\ and\ \citenamefont
  {Saueressig}}]{Benedetti:2009rx}%
  \BibitemOpen
  \bibfield  {author} {\bibinfo {author} {\bibfnamefont {D.}~\bibnamefont
  {Benedetti}}, \bibinfo {author} {\bibfnamefont {P.~F.}\ \bibnamefont
  {Machado}}, \ and\ \bibinfo {author} {\bibfnamefont {F.}~\bibnamefont
  {Saueressig}},\ }\href {\doibase 10.1142/S0217732309031521} {\bibfield
  {journal} {\bibinfo  {journal} {Mod. Phys. Lett.}\ }\textbf {\bibinfo
  {volume} {A24}},\ \bibinfo {pages} {2233} (\bibinfo {year} {2009})},\ \Eprint
  {http://arxiv.org/abs/0901.2984} {arXiv:0901.2984 [hep-th]} \BibitemShut
  {NoStop}%
\bibitem [{\citenamefont {Falls}\ \emph {et~al.}(2013)\citenamefont {Falls},
  \citenamefont {Litim}, \citenamefont {Nikolakopoulos},\ and\ \citenamefont
  {Rahmede}}]{Falls:2013bv}%
  \BibitemOpen
  \bibfield  {author} {\bibinfo {author} {\bibfnamefont {K.}~\bibnamefont
  {Falls}}, \bibinfo {author} {\bibfnamefont {D.}~\bibnamefont {Litim}},
  \bibinfo {author} {\bibfnamefont {K.}~\bibnamefont {Nikolakopoulos}}, \ and\
  \bibinfo {author} {\bibfnamefont {C.}~\bibnamefont {Rahmede}},\ }\href@noop
  {} {\  (\bibinfo {year} {2013})},\ \Eprint {http://arxiv.org/abs/1301.4191}
  {arXiv:1301.4191 [hep-th]} \BibitemShut {NoStop}%
\bibitem [{\citenamefont {Becker}\ and\ \citenamefont
  {Reuter}(2014)}]{Becker:2014qya}%
  \BibitemOpen
  \bibfield  {author} {\bibinfo {author} {\bibfnamefont {D.}~\bibnamefont
  {Becker}}\ and\ \bibinfo {author} {\bibfnamefont {M.}~\bibnamefont
  {Reuter}},\ }\href {\doibase 10.1016/j.aop.2014.07.023} {\bibfield  {journal}
  {\bibinfo  {journal} {Annals Phys.}\ }\textbf {\bibinfo {volume} {350}},\
  \bibinfo {pages} {225} (\bibinfo {year} {2014})},\ \Eprint
  {http://arxiv.org/abs/1404.4537} {arXiv:1404.4537 [hep-th]} \BibitemShut
  {NoStop}%
\bibitem [{\citenamefont {Christiansen}\ \emph {et~al.}(2015)\citenamefont
  {Christiansen}, \citenamefont {Knorr}, \citenamefont {Meibohm}, \citenamefont
  {Pawlowski},\ and\ \citenamefont {Reichert}}]{Christiansen:2015rva}%
  \BibitemOpen
  \bibfield  {author} {\bibinfo {author} {\bibfnamefont {N.}~\bibnamefont
  {Christiansen}}, \bibinfo {author} {\bibfnamefont {B.}~\bibnamefont {Knorr}},
  \bibinfo {author} {\bibfnamefont {J.}~\bibnamefont {Meibohm}}, \bibinfo
  {author} {\bibfnamefont {J.~M.}\ \bibnamefont {Pawlowski}}, \ and\ \bibinfo
  {author} {\bibfnamefont {M.}~\bibnamefont {Reichert}},\ }\href {\doibase
  10.1103/PhysRevD.92.121501} {\bibfield  {journal} {\bibinfo  {journal} {Phys.
  Rev.}\ }\textbf {\bibinfo {volume} {D92}},\ \bibinfo {pages} {121501}
  (\bibinfo {year} {2015})},\ \Eprint {http://arxiv.org/abs/1506.07016}
  {arXiv:1506.07016 [hep-th]} \BibitemShut {NoStop}%
\bibitem [{\citenamefont {Gies}\ \emph {et~al.}(2016)\citenamefont {Gies},
  \citenamefont {Knorr}, \citenamefont {Lippoldt},\ and\ \citenamefont
  {Saueressig}}]{Gies:2016con}%
  \BibitemOpen
  \bibfield  {author} {\bibinfo {author} {\bibfnamefont {H.}~\bibnamefont
  {Gies}}, \bibinfo {author} {\bibfnamefont {B.}~\bibnamefont {Knorr}},
  \bibinfo {author} {\bibfnamefont {S.}~\bibnamefont {Lippoldt}}, \ and\
  \bibinfo {author} {\bibfnamefont {F.}~\bibnamefont {Saueressig}},\ }\href
  {\doibase 10.1103/PhysRevLett.116.211302} {\bibfield  {journal} {\bibinfo
  {journal} {Phys. Rev. Lett.}\ }\textbf {\bibinfo {volume} {116}},\ \bibinfo
  {pages} {211302} (\bibinfo {year} {2016})},\ \Eprint
  {http://arxiv.org/abs/1601.01800} {arXiv:1601.01800 [hep-th]} \BibitemShut
  {NoStop}%
\bibitem [{\citenamefont {Denz}\ \emph {et~al.}(2018)\citenamefont {Denz},
  \citenamefont {Pawlowski},\ and\ \citenamefont {Reichert}}]{Denz:2016qks}%
  \BibitemOpen
  \bibfield  {author} {\bibinfo {author} {\bibfnamefont {T.}~\bibnamefont
  {Denz}}, \bibinfo {author} {\bibfnamefont {J.~M.}\ \bibnamefont {Pawlowski}},
  \ and\ \bibinfo {author} {\bibfnamefont {M.}~\bibnamefont {Reichert}},\
  }\href {\doibase 10.1140/epjc/s10052-018-5806-0} {\bibfield  {journal}
  {\bibinfo  {journal} {Eur. Phys. J.}\ }\textbf {\bibinfo {volume} {C78}},\
  \bibinfo {pages} {336} (\bibinfo {year} {2018})},\ \Eprint
  {http://arxiv.org/abs/1612.07315} {arXiv:1612.07315 [hep-th]} \BibitemShut
  {NoStop}%
\bibitem [{\citenamefont {Eichhorn}\ \emph
  {et~al.}(2019{\natexlab{b}})\citenamefont {Eichhorn}, \citenamefont
  {Lippoldt}, \citenamefont {Pawlowski}, \citenamefont {Reichert},\ and\
  \citenamefont {Schiffer}}]{Eichhorn:2018ydy}%
  \BibitemOpen
  \bibfield  {author} {\bibinfo {author} {\bibfnamefont {A.}~\bibnamefont
  {Eichhorn}}, \bibinfo {author} {\bibfnamefont {S.}~\bibnamefont {Lippoldt}},
  \bibinfo {author} {\bibfnamefont {J.~M.}\ \bibnamefont {Pawlowski}}, \bibinfo
  {author} {\bibfnamefont {M.}~\bibnamefont {Reichert}}, \ and\ \bibinfo
  {author} {\bibfnamefont {M.}~\bibnamefont {Schiffer}},\ }\href {\doibase
  10.1016/j.physletb.2019.01.071} {\bibfield  {journal} {\bibinfo  {journal}
  {Phys. Lett.}\ }\textbf {\bibinfo {volume} {B792}},\ \bibinfo {pages} {310}
  (\bibinfo {year} {2019}{\natexlab{b}})},\ \Eprint
  {http://arxiv.org/abs/1810.02828} {arXiv:1810.02828 [hep-th]} \BibitemShut
  {NoStop}%
\bibitem [{\citenamefont {Falls}\ \emph {et~al.}(2020)\citenamefont {Falls},
  \citenamefont {Ohta},\ and\ \citenamefont {Percacci}}]{Falls:2020qhj}%
  \BibitemOpen
  \bibfield  {author} {\bibinfo {author} {\bibfnamefont {K.}~\bibnamefont
  {Falls}}, \bibinfo {author} {\bibfnamefont {N.}~\bibnamefont {Ohta}}, \ and\
  \bibinfo {author} {\bibfnamefont {R.}~\bibnamefont {Percacci}},\ }\href@noop
  {} {\  (\bibinfo {year} {2020})},\ \Eprint {http://arxiv.org/abs/2004.04126}
  {arXiv:2004.04126 [hep-th]} \BibitemShut {NoStop}%
\bibitem [{\citenamefont {Niedermaier}\ and\ \citenamefont
  {Reuter}(2006)}]{Niedermaier:2006wt}%
  \BibitemOpen
  \bibfield  {author} {\bibinfo {author} {\bibfnamefont {M.}~\bibnamefont
  {Niedermaier}}\ and\ \bibinfo {author} {\bibfnamefont {M.}~\bibnamefont
  {Reuter}},\ }\href {\doibase 10.12942/lrr-2006-5} {\bibfield  {journal}
  {\bibinfo  {journal} {Living Rev.\ Rel.}\ }\textbf {\bibinfo {volume} {9}},\
  \bibinfo {pages} {5} (\bibinfo {year} {2006})}\BibitemShut {NoStop}%
\bibitem [{\citenamefont {Litim}(2011)}]{Litim:2011cp}%
  \BibitemOpen
  \bibfield  {author} {\bibinfo {author} {\bibfnamefont {D.~F.}\ \bibnamefont
  {Litim}},\ }\href {\doibase 10.1098/rsta.2011.0103} {\bibfield  {journal}
  {\bibinfo  {journal} {Phil. Trans. Roy. Soc. Lond. A}\ }\textbf {\bibinfo
  {volume} {369}},\ \bibinfo {pages} {2759} (\bibinfo {year} {2011})},\ \Eprint
  {http://arxiv.org/abs/1102.4624} {arXiv:1102.4624 [hep-th]} \BibitemShut
  {NoStop}%
\bibitem [{\citenamefont {Percacci}(2011)}]{Percacci:2011fr}%
  \BibitemOpen
  \bibfield  {author} {\bibinfo {author} {\bibfnamefont {R.}~\bibnamefont
  {Percacci}},\ }in\ \href@noop {} {\emph {\bibinfo {booktitle} {{Time and
  Matter}}}}\ (\bibinfo {year} {2011})\ pp.\ \bibinfo {pages} {123--142},\
  \Eprint {http://arxiv.org/abs/1110.6389} {arXiv:1110.6389 [hep-th]}
  \BibitemShut {NoStop}%
\bibitem [{\citenamefont {Reuter}\ and\ \citenamefont
  {Saueressig}(2012)}]{Reuter:2012id}%
  \BibitemOpen
  \bibfield  {author} {\bibinfo {author} {\bibfnamefont {M.}~\bibnamefont
  {Reuter}}\ and\ \bibinfo {author} {\bibfnamefont {F.}~\bibnamefont
  {Saueressig}},\ }\href {\doibase 10.1088/1367-2630/14/5/055022} {\bibfield
  {journal} {\bibinfo  {journal} {New J. Phys.}\ }\textbf {\bibinfo {volume}
  {14}},\ \bibinfo {pages} {055022} (\bibinfo {year} {2012})},\ \Eprint
  {http://arxiv.org/abs/1202.2274} {arXiv:1202.2274 [hep-th]} \BibitemShut
  {NoStop}%
\bibitem [{\citenamefont {Ashtekar}\ \emph {et~al.}(2014)\citenamefont
  {Ashtekar}, \citenamefont {Reuter},\ and\ \citenamefont
  {Rovelli}}]{Ashtekar:2014kba}%
  \BibitemOpen
  \bibfield  {author} {\bibinfo {author} {\bibfnamefont {A.}~\bibnamefont
  {Ashtekar}}, \bibinfo {author} {\bibfnamefont {M.}~\bibnamefont {Reuter}}, \
  and\ \bibinfo {author} {\bibfnamefont {C.}~\bibnamefont {Rovelli}},\
  }\href@noop {} {\  (\bibinfo {year} {2014})},\ \Eprint
  {http://arxiv.org/abs/1408.4336} {arXiv:1408.4336 [gr-qc]} \BibitemShut
  {NoStop}%
\bibitem [{\citenamefont {Percacci}(2017)}]{Percacci:2017fkn}%
  \BibitemOpen
  \bibfield  {author} {\bibinfo {author} {\bibfnamefont {R.}~\bibnamefont
  {Percacci}},\ }\href {\doibase 10.1142/10369} {\emph {\bibinfo {title} {{An
  Introduction to Covariant Quantum Gravity and Asymptotic Safety}}}},\
  \bibinfo {series} {{100 Years of General Relativity}}, Vol.~\bibinfo {volume}
  {3}\ (\bibinfo  {publisher} {World Scientific},\ \bibinfo {year}
  {2017})\BibitemShut {NoStop}%
\bibitem [{\citenamefont {Eichhorn}(2018)}]{Eichhorn:2017egq}%
  \BibitemOpen
  \bibfield  {author} {\bibinfo {author} {\bibfnamefont {A.}~\bibnamefont
  {Eichhorn}},\ }\bibfield  {booktitle} {\emph {\bibinfo {booktitle} {{Black
  Holes, Gravitational Waves and Spacetime Singularities Rome, Italy, May 9-12,
  2017}}},\ }\href {\doibase 10.1007/s10701-018-0196-6} {\bibfield  {journal}
  {\bibinfo  {journal} {Found. Phys.}\ }\textbf {\bibinfo {volume} {48}},\
  \bibinfo {pages} {1407} (\bibinfo {year} {2018})},\ \Eprint
  {http://arxiv.org/abs/1709.03696} {arXiv:1709.03696 [gr-qc]} \BibitemShut
  {NoStop}%
\bibitem [{\citenamefont {Pereira}(2019)}]{Pereira:2019dbn}%
  \BibitemOpen
  \bibfield  {author} {\bibinfo {author} {\bibfnamefont {A.~D.}\ \bibnamefont
  {Pereira}},\ }in\ \href@noop {} {\emph {\bibinfo {booktitle} {{Progress and
  Visions in Quantum Theory in View of Gravity}: {Bridging foundations of
  physics and mathematics}}}}\ (\bibinfo {year} {2019})\ \Eprint
  {http://arxiv.org/abs/1904.07042} {arXiv:1904.07042 [gr-qc]} \BibitemShut
  {NoStop}%
\bibitem [{\citenamefont {Reuter}\ and\ \citenamefont
  {Saueressig}(2019)}]{Reuter:2019byg}%
  \BibitemOpen
  \bibfield  {author} {\bibinfo {author} {\bibfnamefont {M.}~\bibnamefont
  {Reuter}}\ and\ \bibinfo {author} {\bibfnamefont {F.}~\bibnamefont
  {Saueressig}},\ }\href
  {https://www.cambridge.org/academic/subjects/physics/theoretical-physics-and-mathematical-physics/quantum-gravity-and-functional-renormalization-group-road-towards-asymptotic-safety?format=HB&isbn=9781107107328}
  {\emph {\bibinfo {title} {{Quantum Gravity and the Functional Renormalization
  Group}}}}\ (\bibinfo  {publisher} {Cambridge University Press},\ \bibinfo
  {year} {2019})\BibitemShut {NoStop}%
\bibitem [{\citenamefont {Eichhorn}(2020)}]{Eichhorn:2020mte}%
  \BibitemOpen
  \bibfield  {author} {\bibinfo {author} {\bibfnamefont {A.}~\bibnamefont
  {Eichhorn}},\ }in\ \href@noop {} {\emph {\bibinfo {booktitle} {{57th
  International School of Subnuclear Physics}: {In Search for the
  Unexpected}}}}\ (\bibinfo {year} {2020})\ \Eprint
  {http://arxiv.org/abs/2003.00044} {arXiv:2003.00044 [gr-qc]} \BibitemShut
  {NoStop}%
\bibitem [{\citenamefont {Reichert}(2020)}]{Reichert:2020mja}%
  \BibitemOpen
  \bibfield  {author} {\bibinfo {author} {\bibfnamefont {M.}~\bibnamefont
  {Reichert}},\ }\href {\doibase 10.22323/1.384.0005} {\bibfield  {journal}
  {\bibinfo  {journal} {PoS}\ }\textbf {\bibinfo {volume} {Modave2019}},\
  \bibinfo {pages} {005} (\bibinfo {year} {2020})}\BibitemShut {NoStop}%
\bibitem [{\citenamefont {Donoghue}(2020)}]{Donoghue:2019clr}%
  \BibitemOpen
  \bibfield  {author} {\bibinfo {author} {\bibfnamefont {J.~F.}\ \bibnamefont
  {Donoghue}},\ }\href {\doibase 10.3389/fphy.2020.00056} {\bibfield  {journal}
  {\bibinfo  {journal} {Front.in Phys.}\ }\textbf {\bibinfo {volume} {8}},\
  \bibinfo {pages} {56} (\bibinfo {year} {2020})},\ \Eprint
  {http://arxiv.org/abs/1911.02967} {arXiv:1911.02967 [hep-th]} \BibitemShut
  {NoStop}%
\bibitem [{\citenamefont {Bonanno}\ \emph {et~al.}(2020)\citenamefont
  {Bonanno}, \citenamefont {Eichhorn}, \citenamefont {Gies}, \citenamefont
  {Pawlowski}, \citenamefont {Percacci}, \citenamefont {Reuter}, \citenamefont
  {Saueressig},\ and\ \citenamefont {Vacca}}]{Bonanno:2020bil}%
  \BibitemOpen
  \bibfield  {author} {\bibinfo {author} {\bibfnamefont {A.}~\bibnamefont
  {Bonanno}}, \bibinfo {author} {\bibfnamefont {A.}~\bibnamefont {Eichhorn}},
  \bibinfo {author} {\bibfnamefont {H.}~\bibnamefont {Gies}}, \bibinfo {author}
  {\bibfnamefont {J.~M.}\ \bibnamefont {Pawlowski}}, \bibinfo {author}
  {\bibfnamefont {R.}~\bibnamefont {Percacci}}, \bibinfo {author}
  {\bibfnamefont {M.}~\bibnamefont {Reuter}}, \bibinfo {author} {\bibfnamefont
  {F.}~\bibnamefont {Saueressig}}, \ and\ \bibinfo {author} {\bibfnamefont
  {G.~P.}\ \bibnamefont {Vacca}},\ }\href@noop {} {\  (\bibinfo {year}
  {2020})},\ \Eprint {http://arxiv.org/abs/2004.06810} {arXiv:2004.06810
  [gr-qc]} \BibitemShut {NoStop}%
\bibitem [{\citenamefont {Eichhorn}\ and\ \citenamefont
  {Gies}(2011)}]{Eichhorn:2011pc}%
  \BibitemOpen
  \bibfield  {author} {\bibinfo {author} {\bibfnamefont {A.}~\bibnamefont
  {Eichhorn}}\ and\ \bibinfo {author} {\bibfnamefont {H.}~\bibnamefont
  {Gies}},\ }\href {\doibase 10.1088/1367-2630/13/12/125012} {\bibfield
  {journal} {\bibinfo  {journal} {New J. Phys.}\ }\textbf {\bibinfo {volume}
  {13}},\ \bibinfo {pages} {125012} (\bibinfo {year} {2011})},\ \Eprint
  {http://arxiv.org/abs/1104.5366} {arXiv:1104.5366 [hep-th]} \BibitemShut
  {NoStop}%
\bibitem [{\citenamefont {Don{\`a}}\ \emph {et~al.}(2014)\citenamefont
  {Don{\`a}}, \citenamefont {Eichhorn},\ and\ \citenamefont
  {Percacci}}]{Dona:2013qba}%
  \BibitemOpen
  \bibfield  {author} {\bibinfo {author} {\bibfnamefont {P.}~\bibnamefont
  {Don{\`a}}}, \bibinfo {author} {\bibfnamefont {A.}~\bibnamefont {Eichhorn}},
  \ and\ \bibinfo {author} {\bibfnamefont {R.}~\bibnamefont {Percacci}},\
  }\href {\doibase 10.1103/PhysRevD.89.084035} {\bibfield  {journal} {\bibinfo
  {journal} {Phys.Rev.}\ }\textbf {\bibinfo {volume} {D89}},\ \bibinfo {pages}
  {084035} (\bibinfo {year} {2014})},\ \Eprint {http://arxiv.org/abs/1311.2898}
  {arXiv:1311.2898 [hep-th]} \BibitemShut {NoStop}%
\bibitem [{\citenamefont {Eichhorn}\ and\ \citenamefont
  {Held}(2017)}]{Eichhorn:2017eht}%
  \BibitemOpen
  \bibfield  {author} {\bibinfo {author} {\bibfnamefont {A.}~\bibnamefont
  {Eichhorn}}\ and\ \bibinfo {author} {\bibfnamefont {A.}~\bibnamefont
  {Held}},\ }\href {\doibase 10.1103/PhysRevD.96.086025} {\bibfield  {journal}
  {\bibinfo  {journal} {Phys. Rev.}\ }\textbf {\bibinfo {volume} {D96}},\
  \bibinfo {pages} {086025} (\bibinfo {year} {2017})},\ \Eprint
  {http://arxiv.org/abs/1705.02342} {arXiv:1705.02342 [gr-qc]} \BibitemShut
  {NoStop}%
\bibitem [{\citenamefont {Meibohm}\ \emph {et~al.}(2016)\citenamefont
  {Meibohm}, \citenamefont {Pawlowski},\ and\ \citenamefont
  {Reichert}}]{Meibohm:2015twa}%
  \BibitemOpen
  \bibfield  {author} {\bibinfo {author} {\bibfnamefont {J.}~\bibnamefont
  {Meibohm}}, \bibinfo {author} {\bibfnamefont {J.~M.}\ \bibnamefont
  {Pawlowski}}, \ and\ \bibinfo {author} {\bibfnamefont {M.}~\bibnamefont
  {Reichert}},\ }\href {\doibase 10.1103/PhysRevD.93.084035} {\bibfield
  {journal} {\bibinfo  {journal} {Phys. Rev.}\ }\textbf {\bibinfo {volume}
  {D93}},\ \bibinfo {pages} {084035} (\bibinfo {year} {2016})},\ \Eprint
  {http://arxiv.org/abs/1510.07018} {arXiv:1510.07018 [hep-th]} \BibitemShut
  {NoStop}%
\bibitem [{\citenamefont {Biemans}\ \emph {et~al.}(2017)\citenamefont
  {Biemans}, \citenamefont {Platania},\ and\ \citenamefont
  {Saueressig}}]{Biemans:2017zca}%
  \BibitemOpen
  \bibfield  {author} {\bibinfo {author} {\bibfnamefont {J.}~\bibnamefont
  {Biemans}}, \bibinfo {author} {\bibfnamefont {A.}~\bibnamefont {Platania}}, \
  and\ \bibinfo {author} {\bibfnamefont {F.}~\bibnamefont {Saueressig}},\
  }\href {\doibase 10.1007/JHEP05(2017)093} {\bibfield  {journal} {\bibinfo
  {journal} {JHEP}\ }\textbf {\bibinfo {volume} {05}},\ \bibinfo {pages} {093}
  (\bibinfo {year} {2017})},\ \Eprint {http://arxiv.org/abs/1702.06539}
  {arXiv:1702.06539 [hep-th]} \BibitemShut {NoStop}%
\bibitem [{\citenamefont {Alkofer}\ and\ \citenamefont
  {Saueressig}(2018)}]{Alkofer:2018fxj}%
  \BibitemOpen
  \bibfield  {author} {\bibinfo {author} {\bibfnamefont {N.}~\bibnamefont
  {Alkofer}}\ and\ \bibinfo {author} {\bibfnamefont {F.}~\bibnamefont
  {Saueressig}},\ }\href {\doibase 10.1016/j.aop.2018.07.017} {\bibfield
  {journal} {\bibinfo  {journal} {Annals Phys.}\ }\textbf {\bibinfo {volume}
  {396}},\ \bibinfo {pages} {173} (\bibinfo {year} {2018})},\ \Eprint
  {http://arxiv.org/abs/1802.00498} {arXiv:1802.00498 [hep-th]} \BibitemShut
  {NoStop}%
\bibitem [{\citenamefont {Harst}\ and\ \citenamefont
  {Reuter}(2011)}]{Harst:2011zx}%
  \BibitemOpen
  \bibfield  {author} {\bibinfo {author} {\bibfnamefont {U.}~\bibnamefont
  {Harst}}\ and\ \bibinfo {author} {\bibfnamefont {M.}~\bibnamefont {Reuter}},\
  }\href {\doibase 10.1007/JHEP05(2011)119} {\bibfield  {journal} {\bibinfo
  {journal} {JHEP}\ }\textbf {\bibinfo {volume} {05}},\ \bibinfo {pages} {119}
  (\bibinfo {year} {2011})},\ \Eprint {http://arxiv.org/abs/1101.6007}
  {arXiv:1101.6007 [hep-th]} \BibitemShut {NoStop}%
\bibitem [{\citenamefont {Eichhorn}\ and\ \citenamefont
  {Held}(2018{\natexlab{a}})}]{Eichhorn:2017ylw}%
  \BibitemOpen
  \bibfield  {author} {\bibinfo {author} {\bibfnamefont {A.}~\bibnamefont
  {Eichhorn}}\ and\ \bibinfo {author} {\bibfnamefont {A.}~\bibnamefont
  {Held}},\ }\href {\doibase 10.1016/j.physletb.2017.12.040} {\bibfield
  {journal} {\bibinfo  {journal} {Phys. Lett.}\ }\textbf {\bibinfo {volume}
  {B777}},\ \bibinfo {pages} {217} (\bibinfo {year} {2018}{\natexlab{a}})},\
  \Eprint {http://arxiv.org/abs/1707.01107} {arXiv:1707.01107 [hep-th]}
  \BibitemShut {NoStop}%
\bibitem [{\citenamefont {Eichhorn}\ and\ \citenamefont
  {Versteegen}(2018)}]{Eichhorn:2017lry}%
  \BibitemOpen
  \bibfield  {author} {\bibinfo {author} {\bibfnamefont {A.}~\bibnamefont
  {Eichhorn}}\ and\ \bibinfo {author} {\bibfnamefont {F.}~\bibnamefont
  {Versteegen}},\ }\href {\doibase 10.1007/JHEP01(2018)030} {\bibfield
  {journal} {\bibinfo  {journal} {JHEP}\ }\textbf {\bibinfo {volume} {01}},\
  \bibinfo {pages} {030} (\bibinfo {year} {2018})},\ \Eprint
  {http://arxiv.org/abs/1709.07252} {arXiv:1709.07252 [hep-th]} \BibitemShut
  {NoStop}%
\bibitem [{\citenamefont {Silveira}\ and\ \citenamefont
  {Zee}(1985)}]{Silveira:1985rk}%
  \BibitemOpen
  \bibfield  {author} {\bibinfo {author} {\bibfnamefont {V.}~\bibnamefont
  {Silveira}}\ and\ \bibinfo {author} {\bibfnamefont {A.}~\bibnamefont {Zee}},\
  }\href {\doibase 10.1016/0370-2693(85)90624-0} {\bibfield  {journal}
  {\bibinfo  {journal} {Phys.\ Lett.\ B}\ }\textbf {\bibinfo {volume} {161}},\
  \bibinfo {pages} {136} (\bibinfo {year} {1985})}\BibitemShut {NoStop}%
\bibitem [{\citenamefont {McDonald}(1994)}]{McDonald:1993ex}%
  \BibitemOpen
  \bibfield  {author} {\bibinfo {author} {\bibfnamefont {J.}~\bibnamefont
  {McDonald}},\ }\href {\doibase 10.1103/PhysRevD.50.3637} {\bibfield
  {journal} {\bibinfo  {journal} {Phys.\ Rev.\ D}\ }\textbf {\bibinfo {volume}
  {50}},\ \bibinfo {pages} {3637} (\bibinfo {year} {1994})},\ \Eprint
  {http://arxiv.org/abs/hep-ph/0702143} {arXiv:hep-ph/0702143} \BibitemShut
  {NoStop}%
\bibitem [{\citenamefont {Burgess}\ \emph {et~al.}(2001)\citenamefont
  {Burgess}, \citenamefont {Pospelov},\ and\ \citenamefont {ter
  Veldhuis}}]{Burgess:2000yq}%
  \BibitemOpen
  \bibfield  {author} {\bibinfo {author} {\bibfnamefont {C.}~\bibnamefont
  {Burgess}}, \bibinfo {author} {\bibfnamefont {M.}~\bibnamefont {Pospelov}}, \
  and\ \bibinfo {author} {\bibfnamefont {T.}~\bibnamefont {ter Veldhuis}},\
  }\href {\doibase 10.1016/S0550-3213(01)00513-2} {\bibfield  {journal}
  {\bibinfo  {journal} {Nucl.\ Phys.\ B}\ }\textbf {\bibinfo {volume} {619}},\
  \bibinfo {pages} {709} (\bibinfo {year} {2001})},\ \Eprint
  {http://arxiv.org/abs/hep-ph/0011335} {arXiv:hep-ph/0011335} \BibitemShut
  {NoStop}%
\bibitem [{\citenamefont {Bento}\ \emph {et~al.}(2001)\citenamefont {Bento},
  \citenamefont {Bertolami},\ and\ \citenamefont {Rosenfeld}}]{Bento:2001yk}%
  \BibitemOpen
  \bibfield  {author} {\bibinfo {author} {\bibfnamefont {M.}~\bibnamefont
  {Bento}}, \bibinfo {author} {\bibfnamefont {O.}~\bibnamefont {Bertolami}}, \
  and\ \bibinfo {author} {\bibfnamefont {R.}~\bibnamefont {Rosenfeld}},\ }\href
  {\doibase 10.1016/S0370-2693(01)01078-4} {\bibfield  {journal} {\bibinfo
  {journal} {Phys.\ Lett.\ B}\ }\textbf {\bibinfo {volume} {518}},\ \bibinfo
  {pages} {276} (\bibinfo {year} {2001})},\ \Eprint
  {http://arxiv.org/abs/hep-ph/0103340} {arXiv:hep-ph/0103340} \BibitemShut
  {NoStop}%
\bibitem [{\citenamefont {McDonald}(2002)}]{McDonald:2001vt}%
  \BibitemOpen
  \bibfield  {author} {\bibinfo {author} {\bibfnamefont {J.}~\bibnamefont
  {McDonald}},\ }\href {\doibase 10.1103/PhysRevLett.88.091304} {\bibfield
  {journal} {\bibinfo  {journal} {Phys.\ Rev.\ Lett.}\ }\textbf {\bibinfo
  {volume} {88}},\ \bibinfo {pages} {091304} (\bibinfo {year} {2002})},\
  \Eprint {http://arxiv.org/abs/hep-ph/0106249} {arXiv:hep-ph/0106249}
  \BibitemShut {NoStop}%
\bibitem [{\citenamefont {Arcadi}\ \emph {et~al.}(2020)\citenamefont {Arcadi},
  \citenamefont {Djouadi},\ and\ \citenamefont {Raidal}}]{Arcadi:2019lka}%
  \BibitemOpen
  \bibfield  {author} {\bibinfo {author} {\bibfnamefont {G.}~\bibnamefont
  {Arcadi}}, \bibinfo {author} {\bibfnamefont {A.}~\bibnamefont {Djouadi}}, \
  and\ \bibinfo {author} {\bibfnamefont {M.}~\bibnamefont {Raidal}},\ }\href
  {\doibase 10.1016/j.physrep.2019.11.003} {\bibfield  {journal} {\bibinfo
  {journal} {Phys. Rept.}\ }\textbf {\bibinfo {volume} {842}},\ \bibinfo
  {pages} {1} (\bibinfo {year} {2020})},\ \Eprint
  {http://arxiv.org/abs/1903.03616} {arXiv:1903.03616 [hep-ph]} \BibitemShut
  {NoStop}%
\bibitem [{\citenamefont {Lopez-Honorez}\ \emph {et~al.}(2012)\citenamefont
  {Lopez-Honorez}, \citenamefont {Schwetz},\ and\ \citenamefont
  {Zupan}}]{LopezHonorez:2012kv}%
  \BibitemOpen
  \bibfield  {author} {\bibinfo {author} {\bibfnamefont {L.}~\bibnamefont
  {Lopez-Honorez}}, \bibinfo {author} {\bibfnamefont {T.}~\bibnamefont
  {Schwetz}}, \ and\ \bibinfo {author} {\bibfnamefont {J.}~\bibnamefont
  {Zupan}},\ }\href {\doibase 10.1016/j.physletb.2012.07.017} {\bibfield
  {journal} {\bibinfo  {journal} {Phys.\ Lett.\ B}\ }\textbf {\bibinfo {volume}
  {716}},\ \bibinfo {pages} {179} (\bibinfo {year} {2012})},\ \Eprint
  {http://arxiv.org/abs/1203.2064} {arXiv:1203.2064 [hep-ph]} \BibitemShut
  {NoStop}%
\bibitem [{\citenamefont {Dupuis}(2016)}]{Dupuis:2016fda}%
  \BibitemOpen
  \bibfield  {author} {\bibinfo {author} {\bibfnamefont {G.}~\bibnamefont
  {Dupuis}},\ }\href {\doibase 10.1007/JHEP07(2016)008} {\bibfield  {journal}
  {\bibinfo  {journal} {JHEP}\ }\textbf {\bibinfo {volume} {07}},\ \bibinfo
  {pages} {008} (\bibinfo {year} {2016})},\ \Eprint
  {http://arxiv.org/abs/1604.04552} {arXiv:1604.04552 [hep-ph]} \BibitemShut
  {NoStop}%
\bibitem [{\citenamefont {Matsumoto}\ \emph {et~al.}(2019)\citenamefont
  {Matsumoto}, \citenamefont {Tsai},\ and\ \citenamefont
  {Tseng}}]{Matsumoto:2018acr}%
  \BibitemOpen
  \bibfield  {author} {\bibinfo {author} {\bibfnamefont {S.}~\bibnamefont
  {Matsumoto}}, \bibinfo {author} {\bibfnamefont {Y.-L.~S.}\ \bibnamefont
  {Tsai}}, \ and\ \bibinfo {author} {\bibfnamefont {P.-Y.}\ \bibnamefont
  {Tseng}},\ }\href {\doibase 10.1007/JHEP07(2019)050} {\bibfield  {journal}
  {\bibinfo  {journal} {JHEP}\ }\textbf {\bibinfo {volume} {07}},\ \bibinfo
  {pages} {050} (\bibinfo {year} {2019})},\ \Eprint
  {http://arxiv.org/abs/1811.03292} {arXiv:1811.03292 [hep-ph]} \BibitemShut
  {NoStop}%
\bibitem [{\citenamefont {Gies}\ and\ \citenamefont
  {Scherer}(2010)}]{Gies:2009hq}%
  \BibitemOpen
  \bibfield  {author} {\bibinfo {author} {\bibfnamefont {H.}~\bibnamefont
  {Gies}}\ and\ \bibinfo {author} {\bibfnamefont {M.~M.}\ \bibnamefont
  {Scherer}},\ }\href {\doibase 10.1140/epjc/s10052-010-1256-z} {\bibfield
  {journal} {\bibinfo  {journal} {Eur. Phys. J.}\ }\textbf {\bibinfo {volume}
  {C66}},\ \bibinfo {pages} {387} (\bibinfo {year} {2010})},\ \Eprint
  {http://arxiv.org/abs/0901.2459} {arXiv:0901.2459 [hep-th]} \BibitemShut
  {NoStop}%
\bibitem [{\citenamefont {Gies}\ \emph {et~al.}(2010)\citenamefont {Gies},
  \citenamefont {Rechenberger},\ and\ \citenamefont {Scherer}}]{Gies:2009sv}%
  \BibitemOpen
  \bibfield  {author} {\bibinfo {author} {\bibfnamefont {H.}~\bibnamefont
  {Gies}}, \bibinfo {author} {\bibfnamefont {S.}~\bibnamefont {Rechenberger}},
  \ and\ \bibinfo {author} {\bibfnamefont {M.~M.}\ \bibnamefont {Scherer}},\
  }\href {\doibase 10.1140/epjc/s10052-010-1257-y} {\bibfield  {journal}
  {\bibinfo  {journal} {Eur. Phys. J.}\ }\textbf {\bibinfo {volume} {C66}},\
  \bibinfo {pages} {403} (\bibinfo {year} {2010})},\ \Eprint
  {http://arxiv.org/abs/0907.0327} {arXiv:0907.0327 [hep-th]} \BibitemShut
  {NoStop}%
\bibitem [{\citenamefont {Vacca}\ and\ \citenamefont
  {Zambelli}(2015)}]{Vacca:2015nta}%
  \BibitemOpen
  \bibfield  {author} {\bibinfo {author} {\bibfnamefont {G.~P.}\ \bibnamefont
  {Vacca}}\ and\ \bibinfo {author} {\bibfnamefont {L.}~\bibnamefont
  {Zambelli}},\ }\href {\doibase 10.1103/PhysRevD.91.125003} {\bibfield
  {journal} {\bibinfo  {journal} {Phys. Rev.}\ }\textbf {\bibinfo {volume}
  {D91}},\ \bibinfo {pages} {125003} (\bibinfo {year} {2015})},\ \Eprint
  {http://arxiv.org/abs/1503.09136} {arXiv:1503.09136 [hep-th]} \BibitemShut
  {NoStop}%
\bibitem [{\citenamefont {Wetterich}(1993)}]{Wetterich:1992yh}%
  \BibitemOpen
  \bibfield  {author} {\bibinfo {author} {\bibfnamefont {C.}~\bibnamefont
  {Wetterich}},\ }\href {\doibase 10.1016/0370-2693(93)90726-X} {\bibfield
  {journal} {\bibinfo  {journal} {Phys. Lett.}\ }\textbf {\bibinfo {volume}
  {B301}},\ \bibinfo {pages} {90} (\bibinfo {year} {1993})},\ \Eprint
  {http://arxiv.org/abs/1710.05815} {arXiv:1710.05815 [hep-th]} \BibitemShut
  {NoStop}%
\bibitem [{\citenamefont {Ellwanger}(1994)}]{Ellwanger:1993mw}%
  \BibitemOpen
  \bibfield  {author} {\bibinfo {author} {\bibfnamefont {U.}~\bibnamefont
  {Ellwanger}},\ }\bibfield  {booktitle} {\emph {\bibinfo {booktitle}
  {{Proceedings, Workshop on Quantum field theoretical aspects of high energy
  physics: Bad Frankenhausen, Germany, September 20-24, 1993}}},\ }\href
  {\doibase 10.1007/BF01555911} {\bibfield  {journal} {\bibinfo  {journal} {Z.
  Phys.}\ }\textbf {\bibinfo {volume} {C62}},\ \bibinfo {pages} {503} (\bibinfo
  {year} {1994})},\ \bibinfo {note} {[,206(1993)]},\ \Eprint
  {http://arxiv.org/abs/hep-ph/9308260} {arXiv:hep-ph/9308260 [hep-ph]}
  \BibitemShut {NoStop}%
\bibitem [{\citenamefont {Morris}(1994)}]{Morris:1993qb}%
  \BibitemOpen
  \bibfield  {author} {\bibinfo {author} {\bibfnamefont {T.~R.}\ \bibnamefont
  {Morris}},\ }\href {\doibase 10.1142/S0217751X94000972} {\bibfield  {journal}
  {\bibinfo  {journal} {Int. J. Mod. Phys.}\ }\textbf {\bibinfo {volume}
  {A9}},\ \bibinfo {pages} {2411} (\bibinfo {year} {1994})},\ \Eprint
  {http://arxiv.org/abs/hep-ph/9308265} {arXiv:hep-ph/9308265} \BibitemShut
  {NoStop}%
\bibitem [{\citenamefont {Pawlowski}(2007)}]{Pawlowski:2005xe}%
  \BibitemOpen
  \bibfield  {author} {\bibinfo {author} {\bibfnamefont {J.~M.}\ \bibnamefont
  {Pawlowski}},\ }\href {\doibase 10.1016/j.aop.2007.01.007} {\bibfield
  {journal} {\bibinfo  {journal} {Annals Phys.}\ }\textbf {\bibinfo {volume}
  {322}},\ \bibinfo {pages} {2831} (\bibinfo {year} {2007})},\ \Eprint
  {http://arxiv.org/abs/hep-th/0512261} {arXiv:hep-th/0512261 [hep-th]}
  \BibitemShut {NoStop}%
\bibitem [{\citenamefont {Gies}(2012)}]{Gies:2006wv}%
  \BibitemOpen
  \bibfield  {author} {\bibinfo {author} {\bibfnamefont {H.}~\bibnamefont
  {Gies}},\ }\href {\doibase 10.1007/978-3-642-27320-9\_6} {\bibfield
  {journal} {\bibinfo  {journal} {Lect.\ Notes Phys.}\ }\textbf {\bibinfo
  {volume} {852}},\ \bibinfo {pages} {287} (\bibinfo {year} {2012})},\ \Eprint
  {http://arxiv.org/abs/hep-ph/0611146} {arXiv:hep-ph/0611146} \BibitemShut
  {NoStop}%
\bibitem [{\citenamefont {Rosten}(2010)}]{Rosten:2010vm}%
  \BibitemOpen
  \bibfield  {author} {\bibinfo {author} {\bibfnamefont {O.~J.}\ \bibnamefont
  {Rosten}},\ }\href@noop {} {\  (\bibinfo {year} {2010})},\ \Eprint
  {http://arxiv.org/abs/1003.1366} {arXiv:1003.1366 [hep-th]} \BibitemShut
  {NoStop}%
\bibitem [{\citenamefont {Vacca}\ and\ \citenamefont
  {Zanusso}(2010)}]{Vacca:2010mj}%
  \BibitemOpen
  \bibfield  {author} {\bibinfo {author} {\bibfnamefont {G.~P.}\ \bibnamefont
  {Vacca}}\ and\ \bibinfo {author} {\bibfnamefont {O.}~\bibnamefont
  {Zanusso}},\ }\href {\doibase 10.1103/PhysRevLett.105.231601} {\bibfield
  {journal} {\bibinfo  {journal} {Phys. Rev. Lett.}\ }\textbf {\bibinfo
  {volume} {105}},\ \bibinfo {pages} {231601} (\bibinfo {year} {2010})},\
  \Eprint {http://arxiv.org/abs/1009.1735} {arXiv:1009.1735 [hep-th]}
  \BibitemShut {NoStop}%
\bibitem [{\citenamefont {Oda}\ and\ \citenamefont
  {Yamada}(2016)}]{Oda:2015sma}%
  \BibitemOpen
  \bibfield  {author} {\bibinfo {author} {\bibfnamefont {K.-y.}\ \bibnamefont
  {Oda}}\ and\ \bibinfo {author} {\bibfnamefont {M.}~\bibnamefont {Yamada}},\
  }\href {\doibase 10.1088/0264-9381/33/12/125011} {\bibfield  {journal}
  {\bibinfo  {journal} {Class. Quant. Grav.}\ }\textbf {\bibinfo {volume}
  {33}},\ \bibinfo {pages} {125011} (\bibinfo {year} {2016})},\ \Eprint
  {http://arxiv.org/abs/1510.03734} {arXiv:1510.03734 [hep-th]} \BibitemShut
  {NoStop}%
\bibitem [{\citenamefont {Eichhorn}\ \emph {et~al.}(2016)\citenamefont
  {Eichhorn}, \citenamefont {Held},\ and\ \citenamefont
  {Pawlowski}}]{Eichhorn:2016esv}%
  \BibitemOpen
  \bibfield  {author} {\bibinfo {author} {\bibfnamefont {A.}~\bibnamefont
  {Eichhorn}}, \bibinfo {author} {\bibfnamefont {A.}~\bibnamefont {Held}}, \
  and\ \bibinfo {author} {\bibfnamefont {J.~M.}\ \bibnamefont {Pawlowski}},\
  }\href {\doibase 10.1103/PhysRevD.94.104027} {\bibfield  {journal} {\bibinfo
  {journal} {Phys. Rev.}\ }\textbf {\bibinfo {volume} {D94}},\ \bibinfo {pages}
  {104027} (\bibinfo {year} {2016})},\ \Eprint
  {http://arxiv.org/abs/1604.02041} {arXiv:1604.02041 [hep-th]} \BibitemShut
  {NoStop}%
\bibitem [{\citenamefont {Hamada}\ and\ \citenamefont
  {Yamada}(2017)}]{Hamada:2017rvn}%
  \BibitemOpen
  \bibfield  {author} {\bibinfo {author} {\bibfnamefont {Y.}~\bibnamefont
  {Hamada}}\ and\ \bibinfo {author} {\bibfnamefont {M.}~\bibnamefont
  {Yamada}},\ }\href {\doibase 10.1007/JHEP08(2017)070} {\bibfield  {journal}
  {\bibinfo  {journal} {JHEP}\ }\textbf {\bibinfo {volume} {08}},\ \bibinfo
  {pages} {070} (\bibinfo {year} {2017})},\ \Eprint
  {http://arxiv.org/abs/1703.09033} {arXiv:1703.09033 [hep-th]} \BibitemShut
  {NoStop}%
\bibitem [{\citenamefont {De~Brito}\ \emph
  {et~al.}(2019{\natexlab{b}})\citenamefont {De~Brito}, \citenamefont {Hamada},
  \citenamefont {Pereira},\ and\ \citenamefont {Yamada}}]{deBrito:2019epw}%
  \BibitemOpen
  \bibfield  {author} {\bibinfo {author} {\bibfnamefont {G.~P.}\ \bibnamefont
  {De~Brito}}, \bibinfo {author} {\bibfnamefont {Y.}~\bibnamefont {Hamada}},
  \bibinfo {author} {\bibfnamefont {A.~D.}\ \bibnamefont {Pereira}}, \ and\
  \bibinfo {author} {\bibfnamefont {M.}~\bibnamefont {Yamada}},\ }\href
  {\doibase 10.1007/JHEP08(2019)142} {\bibfield  {journal} {\bibinfo  {journal}
  {JHEP}\ }\textbf {\bibinfo {volume} {08}},\ \bibinfo {pages} {142} (\bibinfo
  {year} {2019}{\natexlab{b}})},\ \Eprint {http://arxiv.org/abs/1905.11114}
  {arXiv:1905.11114 [hep-th]} \BibitemShut {NoStop}%
\bibitem [{\citenamefont {Buchbinder}\ \emph {et~al.}(1992)\citenamefont
  {Buchbinder}, \citenamefont {Odintsov},\ and\ \citenamefont
  {Shapiro}}]{Buchbinder:1992rb}%
  \BibitemOpen
  \bibfield  {author} {\bibinfo {author} {\bibfnamefont {I.}~\bibnamefont
  {Buchbinder}}, \bibinfo {author} {\bibfnamefont {S.}~\bibnamefont
  {Odintsov}}, \ and\ \bibinfo {author} {\bibfnamefont {I.}~\bibnamefont
  {Shapiro}},\ }\href@noop {} {\emph {\bibinfo {title} {{Effective action in
  quantum gravity}}}}\ (\bibinfo {year} {1992})\BibitemShut {NoStop}%
\bibitem [{\citenamefont {B\"urger}\ \emph {et~al.}(2019)\citenamefont
  {B\"urger}, \citenamefont {Pawlowski}, \citenamefont {Reichert},\ and\
  \citenamefont {Schaefer}}]{Burger:2019upn}%
  \BibitemOpen
  \bibfield  {author} {\bibinfo {author} {\bibfnamefont {B.}~\bibnamefont
  {B\"urger}}, \bibinfo {author} {\bibfnamefont {J.~M.}\ \bibnamefont
  {Pawlowski}}, \bibinfo {author} {\bibfnamefont {M.}~\bibnamefont {Reichert}},
  \ and\ \bibinfo {author} {\bibfnamefont {B.-J.}\ \bibnamefont {Schaefer}},\
  }\href@noop {} {\  (\bibinfo {year} {2019})},\ \Eprint
  {http://arxiv.org/abs/1912.01624} {arXiv:1912.01624 [hep-th]} \BibitemShut
  {NoStop}%
\bibitem [{\citenamefont {Eichhorn}(2012)}]{Eichhorn:2012va}%
  \BibitemOpen
  \bibfield  {author} {\bibinfo {author} {\bibfnamefont {A.}~\bibnamefont
  {Eichhorn}},\ }\href {\doibase 10.1103/PhysRevD.86.105021} {\bibfield
  {journal} {\bibinfo  {journal} {Phys. Rev.}\ }\textbf {\bibinfo {volume}
  {D86}},\ \bibinfo {pages} {105021} (\bibinfo {year} {2012})},\ \Eprint
  {http://arxiv.org/abs/1204.0965} {arXiv:1204.0965 [gr-qc]} \BibitemShut
  {NoStop}%
\bibitem [{\citenamefont {Eichhorn}\ and\ \citenamefont
  {Pauly}(2020)}]{Eichhorn:2020sbo}%
  \BibitemOpen
  \bibfield  {author} {\bibinfo {author} {\bibfnamefont {A.}~\bibnamefont
  {Eichhorn}}\ and\ \bibinfo {author} {\bibfnamefont {M.}~\bibnamefont
  {Pauly}},\ }\href@noop {} {\  (\bibinfo {year} {2020})},\ \Eprint
  {http://arxiv.org/abs/2009.13543} {arXiv:2009.13543 [hep-th]} \BibitemShut
  {NoStop}%
\bibitem [{\citenamefont {Held}(2020)}]{Held:2020kze}%
  \BibitemOpen
  \bibfield  {author} {\bibinfo {author} {\bibfnamefont {A.}~\bibnamefont
  {Held}},\ }\href@noop {} {\  (\bibinfo {year} {2020})},\ \Eprint
  {http://arxiv.org/abs/2003.13642} {arXiv:2003.13642 [hep-th]} \BibitemShut
  {NoStop}%
\bibitem [{\citenamefont {Eichhorn}\ \emph
  {et~al.}(2019{\natexlab{c}})\citenamefont {Eichhorn}, \citenamefont
  {Lippoldt},\ and\ \citenamefont {Schiffer}}]{Eichhorn:2018nda}%
  \BibitemOpen
  \bibfield  {author} {\bibinfo {author} {\bibfnamefont {A.}~\bibnamefont
  {Eichhorn}}, \bibinfo {author} {\bibfnamefont {S.}~\bibnamefont {Lippoldt}},
  \ and\ \bibinfo {author} {\bibfnamefont {M.}~\bibnamefont {Schiffer}},\
  }\href {\doibase 10.1103/PhysRevD.99.086002} {\bibfield  {journal} {\bibinfo
  {journal} {Phys. Rev.}\ }\textbf {\bibinfo {volume} {D99}},\ \bibinfo {pages}
  {086002} (\bibinfo {year} {2019}{\natexlab{c}})},\ \Eprint
  {http://arxiv.org/abs/1812.08782} {arXiv:1812.08782 [hep-th]} \BibitemShut
  {NoStop}%
\bibitem [{\citenamefont {Eichhorn}\ \emph
  {et~al.}(2018{\natexlab{b}})\citenamefont {Eichhorn}, \citenamefont {Labus},
  \citenamefont {Pawlowski},\ and\ \citenamefont
  {Reichert}}]{Eichhorn:2018akn}%
  \BibitemOpen
  \bibfield  {author} {\bibinfo {author} {\bibfnamefont {A.}~\bibnamefont
  {Eichhorn}}, \bibinfo {author} {\bibfnamefont {P.}~\bibnamefont {Labus}},
  \bibinfo {author} {\bibfnamefont {J.~M.}\ \bibnamefont {Pawlowski}}, \ and\
  \bibinfo {author} {\bibfnamefont {M.}~\bibnamefont {Reichert}},\ }\href
  {\doibase 10.21468/SciPostPhys.5.4.031} {\bibfield  {journal} {\bibinfo
  {journal} {SciPost Phys.}\ }\textbf {\bibinfo {volume} {5}},\ \bibinfo
  {pages} {31} (\bibinfo {year} {2018}{\natexlab{b}})},\ \Eprint
  {http://arxiv.org/abs/1804.00012} {arXiv:1804.00012 [hep-th]} \BibitemShut
  {NoStop}%
\bibitem [{\citenamefont {Falls}\ \emph {et~al.}(2016)\citenamefont {Falls},
  \citenamefont {Litim}, \citenamefont {Nikolakopoulos},\ and\ \citenamefont
  {Rahmede}}]{Falls:2014tra}%
  \BibitemOpen
  \bibfield  {author} {\bibinfo {author} {\bibfnamefont {K.}~\bibnamefont
  {Falls}}, \bibinfo {author} {\bibfnamefont {D.~F.}\ \bibnamefont {Litim}},
  \bibinfo {author} {\bibfnamefont {K.}~\bibnamefont {Nikolakopoulos}}, \ and\
  \bibinfo {author} {\bibfnamefont {C.}~\bibnamefont {Rahmede}},\ }\href
  {\doibase 10.1103/PhysRevD.93.104022} {\bibfield  {journal} {\bibinfo
  {journal} {Phys. Rev.}\ }\textbf {\bibinfo {volume} {D93}},\ \bibinfo {pages}
  {104022} (\bibinfo {year} {2016})},\ \Eprint {http://arxiv.org/abs/1410.4815}
  {arXiv:1410.4815 [hep-th]} \BibitemShut {NoStop}%
\bibitem [{\citenamefont {Falls}\ \emph {et~al.}(2018)\citenamefont {Falls},
  \citenamefont {King}, \citenamefont {Litim}, \citenamefont {Nikolakopoulos},\
  and\ \citenamefont {Rahmede}}]{Falls:2017lst}%
  \BibitemOpen
  \bibfield  {author} {\bibinfo {author} {\bibfnamefont {K.}~\bibnamefont
  {Falls}}, \bibinfo {author} {\bibfnamefont {C.~R.}\ \bibnamefont {King}},
  \bibinfo {author} {\bibfnamefont {D.~F.}\ \bibnamefont {Litim}}, \bibinfo
  {author} {\bibfnamefont {K.}~\bibnamefont {Nikolakopoulos}}, \ and\ \bibinfo
  {author} {\bibfnamefont {C.}~\bibnamefont {Rahmede}},\ }\href {\doibase
  10.1103/PhysRevD.97.086006} {\bibfield  {journal} {\bibinfo  {journal} {Phys.
  Rev.}\ }\textbf {\bibinfo {volume} {D97}},\ \bibinfo {pages} {086006}
  (\bibinfo {year} {2018})},\ \Eprint {http://arxiv.org/abs/1801.00162}
  {arXiv:1801.00162 [hep-th]} \BibitemShut {NoStop}%
\bibitem [{\citenamefont {Falls}\ \emph {et~al.}(2019)\citenamefont {Falls},
  \citenamefont {Litim},\ and\ \citenamefont {Schröder}}]{Falls:2018ylp}%
  \BibitemOpen
  \bibfield  {author} {\bibinfo {author} {\bibfnamefont {K.~G.}\ \bibnamefont
  {Falls}}, \bibinfo {author} {\bibfnamefont {D.~F.}\ \bibnamefont {Litim}}, \
  and\ \bibinfo {author} {\bibfnamefont {J.}~\bibnamefont {Schröder}},\ }\href
  {\doibase 10.1103/PhysRevD.99.126015} {\bibfield  {journal} {\bibinfo
  {journal} {Phys. Rev. D}\ }\textbf {\bibinfo {volume} {99}},\ \bibinfo
  {pages} {126015} (\bibinfo {year} {2019})},\ \Eprint
  {http://arxiv.org/abs/1810.08550} {arXiv:1810.08550 [gr-qc]} \BibitemShut
  {NoStop}%
\bibitem [{\citenamefont {Niedermaier}(2009)}]{Niedermaier:2009zz}%
  \BibitemOpen
  \bibfield  {author} {\bibinfo {author} {\bibfnamefont {M.~R.}\ \bibnamefont
  {Niedermaier}},\ }\href {\doibase 10.1103/PhysRevLett.103.101303} {\bibfield
  {journal} {\bibinfo  {journal} {Phys. Rev. Lett.}\ }\textbf {\bibinfo
  {volume} {103}},\ \bibinfo {pages} {101303} (\bibinfo {year}
  {2009})}\BibitemShut {NoStop}%
\bibitem [{\citenamefont {Niedermaier}(2010)}]{Niedermaier:2010zz}%
  \BibitemOpen
  \bibfield  {author} {\bibinfo {author} {\bibfnamefont {M.}~\bibnamefont
  {Niedermaier}},\ }\href {\doibase 10.1016/j.nuclphysb.2010.01.016} {\bibfield
   {journal} {\bibinfo  {journal} {Nucl. Phys.}\ }\textbf {\bibinfo {volume}
  {B833}},\ \bibinfo {pages} {226} (\bibinfo {year} {2010})}\BibitemShut
  {NoStop}%
\bibitem [{\citenamefont {Baek}\ \emph {et~al.}(2012)\citenamefont {Baek},
  \citenamefont {Ko},\ and\ \citenamefont {Park}}]{Baek:2011aa}%
  \BibitemOpen
  \bibfield  {author} {\bibinfo {author} {\bibfnamefont {S.}~\bibnamefont
  {Baek}}, \bibinfo {author} {\bibfnamefont {P.}~\bibnamefont {Ko}}, \ and\
  \bibinfo {author} {\bibfnamefont {W.-I.}\ \bibnamefont {Park}},\ }\href
  {\doibase 10.1007/JHEP02(2012)047} {\bibfield  {journal} {\bibinfo  {journal}
  {JHEP}\ }\textbf {\bibinfo {volume} {02}},\ \bibinfo {pages} {047} (\bibinfo
  {year} {2012})},\ \Eprint {http://arxiv.org/abs/1112.1847} {arXiv:1112.1847
  [hep-ph]} \BibitemShut {NoStop}%
\bibitem [{\citenamefont {Esch}\ \emph {et~al.}(2013)\citenamefont {Esch},
  \citenamefont {Klasen},\ and\ \citenamefont {Yaguna}}]{Esch:2013rta}%
  \BibitemOpen
  \bibfield  {author} {\bibinfo {author} {\bibfnamefont {S.}~\bibnamefont
  {Esch}}, \bibinfo {author} {\bibfnamefont {M.}~\bibnamefont {Klasen}}, \ and\
  \bibinfo {author} {\bibfnamefont {C.~E.}\ \bibnamefont {Yaguna}},\ }\href
  {\doibase 10.1103/PhysRevD.88.075017} {\bibfield  {journal} {\bibinfo
  {journal} {Phys. Rev. D}\ }\textbf {\bibinfo {volume} {88}},\ \bibinfo
  {pages} {075017} (\bibinfo {year} {2013})},\ \Eprint
  {http://arxiv.org/abs/1308.0951} {arXiv:1308.0951 [hep-ph]} \BibitemShut
  {NoStop}%
\bibitem [{\citenamefont {Bagherian}\ \emph {et~al.}(2014)\citenamefont
  {Bagherian}, \citenamefont {Ettefaghi}, \citenamefont {Haghgouyan},\ and\
  \citenamefont {Moazzemi}}]{Bagherian:2014iia}%
  \BibitemOpen
  \bibfield  {author} {\bibinfo {author} {\bibfnamefont {Z.}~\bibnamefont
  {Bagherian}}, \bibinfo {author} {\bibfnamefont {M.~M.}\ \bibnamefont
  {Ettefaghi}}, \bibinfo {author} {\bibfnamefont {Z.}~\bibnamefont
  {Haghgouyan}}, \ and\ \bibinfo {author} {\bibfnamefont {R.}~\bibnamefont
  {Moazzemi}},\ }\href {\doibase 10.1088/1475-7516/2014/10/033} {\bibfield
  {journal} {\bibinfo  {journal} {JCAP}\ }\textbf {\bibinfo {volume} {10}},\
  \bibinfo {pages} {033} (\bibinfo {year} {2014})},\ \Eprint
  {http://arxiv.org/abs/1406.2927} {arXiv:1406.2927 [hep-ph]} \BibitemShut
  {NoStop}%
\bibitem [{\citenamefont {Krnjaic}(2016)}]{Krnjaic:2015mbs}%
  \BibitemOpen
  \bibfield  {author} {\bibinfo {author} {\bibfnamefont {G.}~\bibnamefont
  {Krnjaic}},\ }\href {\doibase 10.1103/PhysRevD.94.073009} {\bibfield
  {journal} {\bibinfo  {journal} {Phys. Rev. D}\ }\textbf {\bibinfo {volume}
  {94}},\ \bibinfo {pages} {073009} (\bibinfo {year} {2016})},\ \Eprint
  {http://arxiv.org/abs/1512.04119} {arXiv:1512.04119 [hep-ph]} \BibitemShut
  {NoStop}%
\bibitem [{\citenamefont {B\'elanger}\ \emph {et~al.}(2018)\citenamefont
  {B\'elanger}, \citenamefont {Boudjema}, \citenamefont {Goudelis},
  \citenamefont {Pukhov},\ and\ \citenamefont {Zaldivar}}]{Belanger:2018ccd}%
  \BibitemOpen
  \bibfield  {author} {\bibinfo {author} {\bibfnamefont {G.}~\bibnamefont
  {B\'elanger}}, \bibinfo {author} {\bibfnamefont {F.}~\bibnamefont
  {Boudjema}}, \bibinfo {author} {\bibfnamefont {A.}~\bibnamefont {Goudelis}},
  \bibinfo {author} {\bibfnamefont {A.}~\bibnamefont {Pukhov}}, \ and\ \bibinfo
  {author} {\bibfnamefont {B.}~\bibnamefont {Zaldivar}},\ }\href {\doibase
  10.1016/j.cpc.2018.04.027} {\bibfield  {journal} {\bibinfo  {journal}
  {Comput. Phys. Commun.}\ }\textbf {\bibinfo {volume} {231}},\ \bibinfo
  {pages} {173} (\bibinfo {year} {2018})},\ \Eprint
  {http://arxiv.org/abs/1801.03509} {arXiv:1801.03509 [hep-ph]} \BibitemShut
  {NoStop}%
\bibitem [{\citenamefont {Semenov}(2009)}]{Semenov:2008jy}%
  \BibitemOpen
  \bibfield  {author} {\bibinfo {author} {\bibfnamefont {A.}~\bibnamefont
  {Semenov}},\ }\href {\doibase 10.1016/j.cpc.2008.10.012} {\bibfield
  {journal} {\bibinfo  {journal} {Comput. Phys. Commun.}\ }\textbf {\bibinfo
  {volume} {180}},\ \bibinfo {pages} {431} (\bibinfo {year} {2009})},\ \Eprint
  {http://arxiv.org/abs/0805.0555} {arXiv:0805.0555 [hep-ph]} \BibitemShut
  {NoStop}%
\bibitem [{\citenamefont {Eichhorn}\ and\ \citenamefont
  {Held}(2018{\natexlab{b}})}]{Eichhorn:2018whv}%
  \BibitemOpen
  \bibfield  {author} {\bibinfo {author} {\bibfnamefont {A.}~\bibnamefont
  {Eichhorn}}\ and\ \bibinfo {author} {\bibfnamefont {A.}~\bibnamefont
  {Held}},\ }\href {\doibase 10.1103/PhysRevLett.121.151302} {\bibfield
  {journal} {\bibinfo  {journal} {Phys. Rev. Lett.}\ }\textbf {\bibinfo
  {volume} {121}},\ \bibinfo {pages} {151302} (\bibinfo {year}
  {2018}{\natexlab{b}})},\ \Eprint {http://arxiv.org/abs/1803.04027}
  {arXiv:1803.04027 [hep-th]} \BibitemShut {NoStop}%
\bibitem [{\citenamefont {Alkofer}\ \emph {et~al.}(2020)\citenamefont
  {Alkofer}, \citenamefont {Eichhorn}, \citenamefont {Held}, \citenamefont
  {Nieto}, \citenamefont {Percacci},\ and\ \citenamefont
  {Schroefl}}]{Alkofer:2020vtb}%
  \BibitemOpen
  \bibfield  {author} {\bibinfo {author} {\bibfnamefont {R.}~\bibnamefont
  {Alkofer}}, \bibinfo {author} {\bibfnamefont {A.}~\bibnamefont {Eichhorn}},
  \bibinfo {author} {\bibfnamefont {A.}~\bibnamefont {Held}}, \bibinfo {author}
  {\bibfnamefont {C.~M.}\ \bibnamefont {Nieto}}, \bibinfo {author}
  {\bibfnamefont {R.}~\bibnamefont {Percacci}}, \ and\ \bibinfo {author}
  {\bibfnamefont {M.}~\bibnamefont {Schroefl}},\ }\href@noop {} {\  (\bibinfo
  {year} {2020})},\ \Eprint {http://arxiv.org/abs/2003.08401} {arXiv:2003.08401
  [hep-ph]} \BibitemShut {NoStop}%
\bibitem [{\citenamefont {Aghanim}\ \emph {et~al.}(2018)\citenamefont {Aghanim}
  \emph {et~al.}}]{Aghanim:2018eyx}%
  \BibitemOpen
  \bibfield  {author} {\bibinfo {author} {\bibfnamefont {N.}~\bibnamefont
  {Aghanim}} \emph {et~al.} (\bibinfo {collaboration} {Planck}),\ }\href@noop
  {} {\  (\bibinfo {year} {2018})},\ \Eprint {http://arxiv.org/abs/1807.06209}
  {arXiv:1807.06209 [astro-ph.CO]} \BibitemShut {NoStop}%
\bibitem [{\citenamefont {O'Connell}\ \emph {et~al.}(2007)\citenamefont
  {O'Connell}, \citenamefont {Ramsey-Musolf},\ and\ \citenamefont
  {Wise}}]{OConnell:2006rsp}%
  \BibitemOpen
  \bibfield  {author} {\bibinfo {author} {\bibfnamefont {D.}~\bibnamefont
  {O'Connell}}, \bibinfo {author} {\bibfnamefont {M.~J.}\ \bibnamefont
  {Ramsey-Musolf}}, \ and\ \bibinfo {author} {\bibfnamefont {M.~B.}\
  \bibnamefont {Wise}},\ }\href {\doibase 10.1103/PhysRevD.75.037701}
  {\bibfield  {journal} {\bibinfo  {journal} {Phys.\ Rev.\ D}\ }\textbf
  {\bibinfo {volume} {75}},\ \bibinfo {pages} {037701} (\bibinfo {year}
  {2007})},\ \Eprint {http://arxiv.org/abs/hep-ph/0611014}
  {arXiv:hep-ph/0611014} \BibitemShut {NoStop}%
\bibitem [{\citenamefont {Aad}\ \emph {et~al.}(2016)\citenamefont {Aad} \emph
  {et~al.}}]{Aad:2015txa}%
  \BibitemOpen
  \bibfield  {author} {\bibinfo {author} {\bibfnamefont {G.}~\bibnamefont
  {Aad}} \emph {et~al.} (\bibinfo {collaboration} {ATLAS}),\ }\href {\doibase
  10.1007/JHEP01(2016)172} {\bibfield  {journal} {\bibinfo  {journal} {JHEP}\
  }\textbf {\bibinfo {volume} {01}},\ \bibinfo {pages} {172} (\bibinfo {year}
  {2016})},\ \Eprint {http://arxiv.org/abs/1508.07869} {arXiv:1508.07869
  [hep-ex]} \BibitemShut {NoStop}%
\bibitem [{\citenamefont {Aad}\ \emph {et~al.}(2015)\citenamefont {Aad} \emph
  {et~al.}}]{Aad:2015pla}%
  \BibitemOpen
  \bibfield  {author} {\bibinfo {author} {\bibfnamefont {G.}~\bibnamefont
  {Aad}} \emph {et~al.} (\bibinfo {collaboration} {ATLAS}),\ }\href {\doibase
  10.1007/JHEP11(2015)206} {\bibfield  {journal} {\bibinfo  {journal} {JHEP}\
  }\textbf {\bibinfo {volume} {11}},\ \bibinfo {pages} {206} (\bibinfo {year}
  {2015})},\ \Eprint {http://arxiv.org/abs/1509.00672} {arXiv:1509.00672
  [hep-ex]} \BibitemShut {NoStop}%
\bibitem [{\citenamefont {Khachatryan}\ \emph {et~al.}(2017)\citenamefont
  {Khachatryan} \emph {et~al.}}]{Khachatryan:2016whc}%
  \BibitemOpen
  \bibfield  {author} {\bibinfo {author} {\bibfnamefont {V.}~\bibnamefont
  {Khachatryan}} \emph {et~al.} (\bibinfo {collaboration} {CMS}),\ }\href
  {\doibase 10.1007/JHEP02(2017)135} {\bibfield  {journal} {\bibinfo  {journal}
  {JHEP}\ }\textbf {\bibinfo {volume} {02}},\ \bibinfo {pages} {135} (\bibinfo
  {year} {2017})},\ \Eprint {http://arxiv.org/abs/1610.09218} {arXiv:1610.09218
  [hep-ex]} \BibitemShut {NoStop}%
\bibitem [{\citenamefont {Sirunyan}\ \emph {et~al.}(2019)\citenamefont
  {Sirunyan} \emph {et~al.}}]{Sirunyan:2018owy}%
  \BibitemOpen
  \bibfield  {author} {\bibinfo {author} {\bibfnamefont {A.~M.}\ \bibnamefont
  {Sirunyan}} \emph {et~al.} (\bibinfo {collaboration} {CMS}),\ }\href
  {\doibase 10.1016/j.physletb.2019.04.025} {\bibfield  {journal} {\bibinfo
  {journal} {Phys. Lett. B}\ }\textbf {\bibinfo {volume} {793}},\ \bibinfo
  {pages} {520} (\bibinfo {year} {2019})},\ \Eprint
  {http://arxiv.org/abs/1809.05937} {arXiv:1809.05937 [hep-ex]} \BibitemShut
  {NoStop}%
\bibitem [{\citenamefont {Figueroa}\ \emph {et~al.}(2018)\citenamefont
  {Figueroa}, \citenamefont {Rajantie},\ and\ \citenamefont
  {Torrenti}}]{Figueroa:2017slm}%
  \BibitemOpen
  \bibfield  {author} {\bibinfo {author} {\bibfnamefont {D.~G.}\ \bibnamefont
  {Figueroa}}, \bibinfo {author} {\bibfnamefont {A.}~\bibnamefont {Rajantie}},
  \ and\ \bibinfo {author} {\bibfnamefont {F.}~\bibnamefont {Torrenti}},\
  }\href {\doibase 10.1103/PhysRevD.98.023532} {\bibfield  {journal} {\bibinfo
  {journal} {Phys. Rev. D}\ }\textbf {\bibinfo {volume} {98}},\ \bibinfo
  {pages} {023532} (\bibinfo {year} {2018})},\ \Eprint
  {http://arxiv.org/abs/1709.00398} {arXiv:1709.00398 [astro-ph.CO]}
  \BibitemShut {NoStop}%
\bibitem [{\citenamefont {Markkanen}\ \emph {et~al.}(2018)\citenamefont
  {Markkanen}, \citenamefont {Rajantie},\ and\ \citenamefont
  {Stopyra}}]{Markkanen:2018pdo}%
  \BibitemOpen
  \bibfield  {author} {\bibinfo {author} {\bibfnamefont {T.}~\bibnamefont
  {Markkanen}}, \bibinfo {author} {\bibfnamefont {A.}~\bibnamefont {Rajantie}},
  \ and\ \bibinfo {author} {\bibfnamefont {S.}~\bibnamefont {Stopyra}},\ }\href
  {\doibase 10.3389/fspas.2018.00040} {\bibfield  {journal} {\bibinfo
  {journal} {Front. Astron. Space Sci.}\ }\textbf {\bibinfo {volume} {5}},\
  \bibinfo {pages} {40} (\bibinfo {year} {2018})},\ \Eprint
  {http://arxiv.org/abs/1809.06923} {arXiv:1809.06923 [astro-ph.CO]}
  \BibitemShut {NoStop}%
\bibitem [{\citenamefont {Cline}\ \emph {et~al.}(2013)\citenamefont {Cline},
  \citenamefont {Kainulainen}, \citenamefont {Scott},\ and\ \citenamefont
  {Weniger}}]{Cline:2013gha}%
  \BibitemOpen
  \bibfield  {author} {\bibinfo {author} {\bibfnamefont {J.~M.}\ \bibnamefont
  {Cline}}, \bibinfo {author} {\bibfnamefont {K.}~\bibnamefont {Kainulainen}},
  \bibinfo {author} {\bibfnamefont {P.}~\bibnamefont {Scott}}, \ and\ \bibinfo
  {author} {\bibfnamefont {C.}~\bibnamefont {Weniger}},\ }\href {\doibase
  10.1103/PhysRevD.88.055025} {\bibfield  {journal} {\bibinfo  {journal} {Phys.
  Rev. D}\ }\textbf {\bibinfo {volume} {88}},\ \bibinfo {pages} {055025}
  (\bibinfo {year} {2013})},\ \bibinfo {note} {[Erratum: Phys.Rev.D 92, 039906
  (2015)]},\ \Eprint {http://arxiv.org/abs/1306.4710} {arXiv:1306.4710
  [hep-ph]} \BibitemShut {NoStop}%
\bibitem [{\citenamefont {Beniwal}\ \emph {et~al.}(2016)\citenamefont
  {Beniwal}, \citenamefont {Rajec}, \citenamefont {Savage}, \citenamefont
  {Scott}, \citenamefont {Weniger}, \citenamefont {White},\ and\ \citenamefont
  {Williams}}]{Beniwal:2015sdl}%
  \BibitemOpen
  \bibfield  {author} {\bibinfo {author} {\bibfnamefont {A.}~\bibnamefont
  {Beniwal}}, \bibinfo {author} {\bibfnamefont {F.}~\bibnamefont {Rajec}},
  \bibinfo {author} {\bibfnamefont {C.}~\bibnamefont {Savage}}, \bibinfo
  {author} {\bibfnamefont {P.}~\bibnamefont {Scott}}, \bibinfo {author}
  {\bibfnamefont {C.}~\bibnamefont {Weniger}}, \bibinfo {author} {\bibfnamefont
  {M.}~\bibnamefont {White}}, \ and\ \bibinfo {author} {\bibfnamefont {A.~G.}\
  \bibnamefont {Williams}},\ }\href {\doibase 10.1103/PhysRevD.93.115016}
  {\bibfield  {journal} {\bibinfo  {journal} {Phys. Rev. D}\ }\textbf {\bibinfo
  {volume} {93}},\ \bibinfo {pages} {115016} (\bibinfo {year} {2016})},\
  \Eprint {http://arxiv.org/abs/1512.06458} {arXiv:1512.06458 [hep-ph]}
  \BibitemShut {NoStop}%
\bibitem [{\citenamefont {Gaskins}(2016)}]{Gaskins:2016cha}%
  \BibitemOpen
  \bibfield  {author} {\bibinfo {author} {\bibfnamefont {J.~M.}\ \bibnamefont
  {Gaskins}},\ }\href {\doibase 10.1080/00107514.2016.1175160} {\bibfield
  {journal} {\bibinfo  {journal} {Contemp. Phys.}\ }\textbf {\bibinfo {volume}
  {57}},\ \bibinfo {pages} {496} (\bibinfo {year} {2016})},\ \Eprint
  {http://arxiv.org/abs/1604.00014} {arXiv:1604.00014 [astro-ph.HE]}
  \BibitemShut {NoStop}%
\bibitem [{\citenamefont {Arcadi}\ \emph {et~al.}(2018)\citenamefont {Arcadi},
  \citenamefont {Dutra}, \citenamefont {Ghosh}, \citenamefont {Lindner},
  \citenamefont {Mambrini}, \citenamefont {Pierre}, \citenamefont {Profumo},\
  and\ \citenamefont {Queiroz}}]{Arcadi:2017kky}%
  \BibitemOpen
  \bibfield  {author} {\bibinfo {author} {\bibfnamefont {G.}~\bibnamefont
  {Arcadi}}, \bibinfo {author} {\bibfnamefont {M.}~\bibnamefont {Dutra}},
  \bibinfo {author} {\bibfnamefont {P.}~\bibnamefont {Ghosh}}, \bibinfo
  {author} {\bibfnamefont {M.}~\bibnamefont {Lindner}}, \bibinfo {author}
  {\bibfnamefont {Y.}~\bibnamefont {Mambrini}}, \bibinfo {author}
  {\bibfnamefont {M.}~\bibnamefont {Pierre}}, \bibinfo {author} {\bibfnamefont
  {S.}~\bibnamefont {Profumo}}, \ and\ \bibinfo {author} {\bibfnamefont
  {F.~S.}\ \bibnamefont {Queiroz}},\ }\href {\doibase
  10.1140/epjc/s10052-018-5662-y} {\bibfield  {journal} {\bibinfo  {journal}
  {Eur. Phys. J. C}\ }\textbf {\bibinfo {volume} {78}},\ \bibinfo {pages} {203}
  (\bibinfo {year} {2018})},\ \Eprint {http://arxiv.org/abs/1703.07364}
  {arXiv:1703.07364 [hep-ph]} \BibitemShut {NoStop}%
\bibitem [{\citenamefont {Roszkowski}\ \emph {et~al.}(2018)\citenamefont
  {Roszkowski}, \citenamefont {Sessolo},\ and\ \citenamefont
  {Trojanowski}}]{Roszkowski:2017nbc}%
  \BibitemOpen
  \bibfield  {author} {\bibinfo {author} {\bibfnamefont {L.}~\bibnamefont
  {Roszkowski}}, \bibinfo {author} {\bibfnamefont {E.~M.}\ \bibnamefont
  {Sessolo}}, \ and\ \bibinfo {author} {\bibfnamefont {S.}~\bibnamefont
  {Trojanowski}},\ }\href {\doibase 10.1088/1361-6633/aab913} {\bibfield
  {journal} {\bibinfo  {journal} {Rept. Prog. Phys.}\ }\textbf {\bibinfo
  {volume} {81}},\ \bibinfo {pages} {066201} (\bibinfo {year} {2018})},\
  \Eprint {http://arxiv.org/abs/1707.06277} {arXiv:1707.06277 [hep-ph]}
  \BibitemShut {NoStop}%
\bibitem [{\citenamefont {Athron}\ \emph {et~al.}(2019)\citenamefont {Athron}
  \emph {et~al.}}]{Athron:2018hpc}%
  \BibitemOpen
  \bibfield  {author} {\bibinfo {author} {\bibfnamefont {P.}~\bibnamefont
  {Athron}} \emph {et~al.} (\bibinfo {collaboration} {GAMBIT}),\ }\href
  {\doibase 10.1140/epjc/s10052-018-6513-6} {\bibfield  {journal} {\bibinfo
  {journal} {Eur. Phys. J. C}\ }\textbf {\bibinfo {volume} {79}},\ \bibinfo
  {pages} {38} (\bibinfo {year} {2019})},\ \Eprint
  {http://arxiv.org/abs/1808.10465} {arXiv:1808.10465 [hep-ph]} \BibitemShut
  {NoStop}%
\bibitem [{\citenamefont {Aprile}\ \emph {et~al.}(2016)\citenamefont {Aprile}
  \emph {et~al.}}]{Aprile:2015uzo}%
  \BibitemOpen
  \bibfield  {author} {\bibinfo {author} {\bibfnamefont {E.}~\bibnamefont
  {Aprile}} \emph {et~al.} (\bibinfo {collaboration} {XENON}),\ }\href
  {\doibase 10.1088/1475-7516/2016/04/027} {\bibfield  {journal} {\bibinfo
  {journal} {JCAP}\ }\textbf {\bibinfo {volume} {04}},\ \bibinfo {pages} {027}
  (\bibinfo {year} {2016})},\ \Eprint {http://arxiv.org/abs/1512.07501}
  {arXiv:1512.07501 [physics.ins-det]} \BibitemShut {NoStop}%
\bibitem [{\citenamefont {Aalbers}\ \emph {et~al.}(2016)\citenamefont {Aalbers}
  \emph {et~al.}}]{Aalbers:2016jon}%
  \BibitemOpen
  \bibfield  {author} {\bibinfo {author} {\bibfnamefont {J.}~\bibnamefont
  {Aalbers}} \emph {et~al.} (\bibinfo {collaboration} {DARWIN}),\ }\href
  {\doibase 10.1088/1475-7516/2016/11/017} {\bibfield  {journal} {\bibinfo
  {journal} {JCAP}\ }\textbf {\bibinfo {volume} {11}},\ \bibinfo {pages} {017}
  (\bibinfo {year} {2016})},\ \Eprint {http://arxiv.org/abs/1606.07001}
  {arXiv:1606.07001 [astro-ph.IM]} \BibitemShut {NoStop}%
\bibitem [{\citenamefont {Agnese}\ \emph {et~al.}(2017)\citenamefont {Agnese}
  \emph {et~al.}}]{Agnese:2016cpb}%
  \BibitemOpen
  \bibfield  {author} {\bibinfo {author} {\bibfnamefont {R.}~\bibnamefont
  {Agnese}} \emph {et~al.} (\bibinfo {collaboration} {SuperCDMS}),\ }\href
  {\doibase 10.1103/PhysRevD.95.082002} {\bibfield  {journal} {\bibinfo
  {journal} {Phys. Rev. D}\ }\textbf {\bibinfo {volume} {95}},\ \bibinfo
  {pages} {082002} (\bibinfo {year} {2017})},\ \Eprint
  {http://arxiv.org/abs/1610.00006} {arXiv:1610.00006 [physics.ins-det]}
  \BibitemShut {NoStop}%
\bibitem [{\citenamefont {Akerib}\ \emph {et~al.}(2020)\citenamefont {Akerib}
  \emph {et~al.}}]{Akerib:2018lyp}%
  \BibitemOpen
  \bibfield  {author} {\bibinfo {author} {\bibfnamefont {D.}~\bibnamefont
  {Akerib}} \emph {et~al.} (\bibinfo {collaboration} {LUX-ZEPLIN}),\ }\href
  {\doibase 10.1103/PhysRevD.101.052002} {\bibfield  {journal} {\bibinfo
  {journal} {Phys. Rev. D}\ }\textbf {\bibinfo {volume} {101}},\ \bibinfo
  {pages} {052002} (\bibinfo {year} {2020})},\ \Eprint
  {http://arxiv.org/abs/1802.06039} {arXiv:1802.06039 [astro-ph.IM]}
  \BibitemShut {NoStop}%
\bibitem [{\citenamefont {Zhang}\ \emph {et~al.}(2019)\citenamefont {Zhang}
  \emph {et~al.}}]{Zhang:2018xdp}%
  \BibitemOpen
  \bibfield  {author} {\bibinfo {author} {\bibfnamefont {H.}~\bibnamefont
  {Zhang}} \emph {et~al.} (\bibinfo {collaboration} {PandaX}),\ }\href
  {\doibase 10.1007/s11433-018-9259-0} {\bibfield  {journal} {\bibinfo
  {journal} {Sci. China Phys. Mech. Astron.}\ }\textbf {\bibinfo {volume}
  {62}},\ \bibinfo {pages} {31011} (\bibinfo {year} {2019})},\ \Eprint
  {http://arxiv.org/abs/1806.02229} {arXiv:1806.02229 [physics.ins-det]}
  \BibitemShut {NoStop}%
\end{thebibliography}%

\end{document}